\documentclass[aps,prd,twocolumn,superscriptaddress,showpacs,preprintnumbers,amsmath,amssymb]{revtex4}

\usepackage{graphicx} 
\usepackage{dcolumn} 
\usepackage{gensymb}

\graphicspath{{ps}}

\begin{document}

\preprint{\vbox{ \hbox{   }
	          	\hbox{Belle Preprint 2017-16}
                         \hbox{KEK Preprint 2017-17}
}}

\title{ \quad\\[1.0cm] Study of \boldmath$\eta$ and dipion transitions in \boldmath$\Upsilon(4S)$ decays to lower bottomonia}

\noaffiliation
\affiliation{University of the Basque Country UPV/EHU, 48080 Bilbao}
\affiliation{Beihang University, Beijing 100191}
\affiliation{Budker Institute of Nuclear Physics SB RAS, Novosibirsk 630090}
\affiliation{Faculty of Mathematics and Physics, Charles University, 121 16 Prague}
\affiliation{Chonnam National University, Kwangju 660-701}
\affiliation{University of Cincinnati, Cincinnati, Ohio 45221}
\affiliation{Deutsches Elektronen--Synchrotron, 22607 Hamburg}
\affiliation{University of Florida, Gainesville, Florida 32611}
\affiliation{Justus-Liebig-Universit\"at Gie\ss{}en, 35392 Gie\ss{}en}
\affiliation{SOKENDAI (The Graduate University for Advanced Studies), Hayama 240-0193}
\affiliation{Gyeongsang National University, Chinju 660-701}
\affiliation{Hanyang University, Seoul 133-791}
\affiliation{University of Hawaii, Honolulu, Hawaii 96822}
\affiliation{High Energy Accelerator Research Organization (KEK), Tsukuba 305-0801}
\affiliation{J-PARC Branch, KEK Theory Center, High Energy Accelerator Research Organization (KEK), Tsukuba 305-0801}
\affiliation{IKERBASQUE, Basque Foundation for Science, 48013 Bilbao}
\affiliation{Indian Institute of Science Education and Research Mohali, SAS Nagar, 140306}
\affiliation{Indian Institute of Technology Bhubaneswar, Satya Nagar 751007}
\affiliation{Indian Institute of Technology Guwahati, Assam 781039}
\affiliation{Indian Institute of Technology Madras, Chennai 600036}
\affiliation{Indiana University, Bloomington, Indiana 47408}
\affiliation{Institute of High Energy Physics, Chinese Academy of Sciences, Beijing 100049}
\affiliation{Institute of High Energy Physics, Vienna 1050}
\affiliation{Institute for High Energy Physics, Protvino 142281}
\affiliation{INFN - Sezione di Napoli, 80126 Napoli}
\affiliation{INFN - Sezione di Torino, 10125 Torino}
\affiliation{Advanced Science Research Center, Japan Atomic Energy Agency, Naka 319-1195}
\affiliation{J. Stefan Institute, 1000 Ljubljana}
\affiliation{Kanagawa University, Yokohama 221-8686}
\affiliation{Institut f\"ur Experimentelle Kernphysik, Karlsruher Institut f\"ur Technologie, 76131 Karlsruhe}
\affiliation{Kennesaw State University, Kennesaw, Georgia 30144}
\affiliation{King Abdulaziz City for Science and Technology, Riyadh 11442}
\affiliation{Department of Physics, Faculty of Science, King Abdulaziz University, Jeddah 21589}
\affiliation{Korea Institute of Science and Technology Information, Daejeon 305-806}
\affiliation{Korea University, Seoul 136-713}
\affiliation{Kyoto University, Kyoto 606-8502}
\affiliation{Kyungpook National University, Daegu 702-701}
\affiliation{\'Ecole Polytechnique F\'ed\'erale de Lausanne (EPFL), Lausanne 1015}
\affiliation{P.N. Lebedev Physical Institute of the Russian Academy of Sciences, Moscow 119991}
\affiliation{Faculty of Mathematics and Physics, University of Ljubljana, 1000 Ljubljana}
\affiliation{Ludwig Maximilians University, 80539 Munich}
\affiliation{Luther College, Decorah, Iowa 52101}
\affiliation{University of Maribor, 2000 Maribor}
\affiliation{Max-Planck-Institut f\"ur Physik, 80805 M\"unchen}
\affiliation{School of Physics, University of Melbourne, Victoria 3010}
\affiliation{University of Miyazaki, Miyazaki 889-2192}
\affiliation{Moscow Physical Engineering Institute, Moscow 115409}
\affiliation{Moscow Institute of Physics and Technology, Moscow Region 141700}
\affiliation{Graduate School of Science, Nagoya University, Nagoya 464-8602}
\affiliation{Kobayashi-Maskawa Institute, Nagoya University, Nagoya 464-8602}
\affiliation{Nara Women's University, Nara 630-8506}
\affiliation{National Central University, Chung-li 32054}
\affiliation{National United University, Miao Li 36003}
\affiliation{Department of Physics, National Taiwan University, Taipei 10617}
\affiliation{H. Niewodniczanski Institute of Nuclear Physics, Krakow 31-342}
\affiliation{Nippon Dental University, Niigata 951-8580}
\affiliation{Niigata University, Niigata 950-2181}
\affiliation{Novosibirsk State University, Novosibirsk 630090}
\affiliation{Osaka City University, Osaka 558-8585}
\affiliation{Pacific Northwest National Laboratory, Richland, Washington 99352}
\affiliation{University of Pittsburgh, Pittsburgh, Pennsylvania 15260}
\affiliation{Theoretical Research Division, Nishina Center, RIKEN, Saitama 351-0198}
\affiliation{University of Science and Technology of China, Hefei 230026}
\affiliation{Showa Pharmaceutical University, Tokyo 194-8543}
\affiliation{Soongsil University, Seoul 156-743}
\affiliation{Stefan Meyer Institute for Subatomic Physics, Vienna 1090}
\affiliation{Sungkyunkwan University, Suwon 440-746}
\affiliation{School of Physics, University of Sydney, New South Wales 2006}
\affiliation{Department of Physics, Faculty of Science, University of Tabuk, Tabuk 71451}
\affiliation{Excellence Cluster Universe, Technische Universit\"at M\"unchen, 85748 Garching}
\affiliation{Department of Physics, Technische Universit\"at M\"unchen, 85748 Garching}
\affiliation{Toho University, Funabashi 274-8510}
\affiliation{Department of Physics, Tohoku University, Sendai 980-8578}
\affiliation{Earthquake Research Institute, University of Tokyo, Tokyo 113-0032}
\affiliation{Department of Physics, University of Tokyo, Tokyo 113-0033}
\affiliation{Tokyo Institute of Technology, Tokyo 152-8550}
\affiliation{Tokyo Metropolitan University, Tokyo 192-0397}
\affiliation{University of Torino, 10124 Torino}
\affiliation{Virginia Polytechnic Institute and State University, Blacksburg, Virginia 24061}
\affiliation{Wayne State University, Detroit, Michigan 48202}
\affiliation{Yamagata University, Yamagata 990-8560}
\affiliation{Yonsei University, Seoul 120-749}
  \author{E.~Guido}\affiliation{INFN - Sezione di Torino, 10125 Torino} 
  \author{R.~Mussa}\affiliation{INFN - Sezione di Torino, 10125 Torino} 
  \author{U.~Tamponi}\affiliation{INFN - Sezione di Torino, 10125 Torino}\affiliation{University of Torino, 10124 Torino} 
  \author{I.~Adachi}\affiliation{High Energy Accelerator Research Organization (KEK), Tsukuba 305-0801}\affiliation{SOKENDAI (The Graduate University for Advanced Studies), Hayama 240-0193} 
  \author{H.~Aihara}\affiliation{Department of Physics, University of Tokyo, Tokyo 113-0033} 
  \author{S.~Al~Said}\affiliation{Department of Physics, Faculty of Science, University of Tabuk, Tabuk 71451}\affiliation{Department of Physics, Faculty of Science, King Abdulaziz University, Jeddah 21589} 
  \author{D.~M.~Asner}\affiliation{Pacific Northwest National Laboratory, Richland, Washington 99352} 
  \author{V.~Aulchenko}\affiliation{Budker Institute of Nuclear Physics SB RAS, Novosibirsk 630090}\affiliation{Novosibirsk State University, Novosibirsk 630090} 
  \author{T.~Aushev}\affiliation{Moscow Institute of Physics and Technology, Moscow Region 141700} 
  \author{R.~Ayad}\affiliation{Department of Physics, Faculty of Science, University of Tabuk, Tabuk 71451} 
% ...
\collaboration{The Belle Collaboration}

\noaffiliation

\begin{abstract}
  We study hadronic transitions between bottomonium states using 496  fb$^{-1}$
   data collected at the $\Upsilon(4S)$ resonance with the Belle detector at the KEKB asymmetric energy $e^{+}e^{-}$ collider. 
  We measure: ${\cal B}(\Upsilon(4S)\to\pi^+\pi^-\Upsilon(1S))=(8.2\pm 0.5 {\rm(stat.)} \pm 0.4 {\rm(syst.)})\times10^{-5}$, ${\cal B}(\Upsilon(4S)\to\pi^+\pi^-\Upsilon(2S))=(7.9\pm 1.0 {\rm(stat.)} \pm 0.4 {\rm(syst.)})\times10^{-5}$, and ${\cal B}(\Upsilon(4S)\to\eta\Upsilon(1S))=(1.70\pm 0.23 {\rm(stat.)} \pm 0.08 {\rm(syst.)})\times10^{-4}$. We measure the ratio  of branching fractions ${\cal R} = {\cal B}(\Upsilon(4S)\to\eta\Upsilon(1S))/{\cal B}(\Upsilon(4S)\to\pi^+\pi^-\Upsilon(1S)) = 2.07\pm 0.30 {\rm(stat.)} \pm 0.11 {\rm(syst.)}$. We search for the decay $\Upsilon(1^3D_{1,2})\to\eta\Upsilon(1S)$, but do not find significant evidence for such a transition. We also measure the initial state radiation production cross sections of the $\Upsilon(2S,3S)$ resonances and we find values compatible with the expected ones. 
 Finally, the analysis of the $\Upsilon(4S)\to\pi^+\pi^-\Upsilon(1S)$ events shows indications for a resonant contribution due to the $f_0(980)$ meson.
\end{abstract}

\pacs{14.40.Pq,13.25.Gv}

\maketitle

\tighten

{\renewcommand{\thefootnote}{\fnsymbol{footnote}}}
\setcounter{footnote}{0}


\section{Introduction}\label{sec:intro}
Recently, hadronic transitions via an $\eta$ meson or two pions between bound states of bottomonium have been recently intensively studied, for instance in~\cite{ref:BaBar4S,ref:Belle4Sdipion,ref:Belle5Sdipion,ref:BaBar3S2S,ref:Belle4Seta}, often with unexpected results. The QCD multipole expansion model~\cite{ref:QCDME} can be used generally to describe hadronic transition between the lower mass bottomonium levels, while its predictions fail when considering bottomonia above the $B\bar{B}$ threshold. In particular, the transitions between bottomonium states via an $\eta$ meson are predicted, for example in~\cite{ref:QCDME,ref:Voloshin,ref:Simonov}, to be highly suppressed, since they require a spin flip of the heavy quark. Among the most unexpected experimental measurements, the BaBar collaboration found an enhancement of the transition $\Upsilon(4S)\to\eta\Upsilon(1S)$ with respect to the transition via a dipion~\cite{ref:BaBar4S}. Also, the Belle collaboration observed the transition $\Upsilon(4S)\to\eta h_b(1P)$ as the non-$B\bar{B}$ transition of the $\Upsilon(4S)$ with the highest branching fraction~\cite{ref:Belle4Seta}. This unsettled picture could be made clearer by the precise measurement of the transitions from the $\Upsilon(4S)$ to lower-mass $\Upsilon$ states via an $\eta$ meson or a dipion, and also by the search for other possible transitions between bottomonia via an $\eta$ meson.

In this paper, we study the transitions $\Upsilon(4S)\to\pi^+\pi^-\Upsilon(nS)$ with $n=1,2$ hereinafter, and $\Upsilon(4S)\to\eta\Upsilon(1S)$, by reconstructing the $\Upsilon(nS)$ mesons via their leptonic decay to two muons. The $\eta$ meson is reconstructed via its decay to $\pi^+\pi^-\pi^0$, with the $\pi^0$ meson reconstructed as two photons. The decay $\eta\to\gamma\gamma$ is not considered in this paper since the corresponding final state has a limited statistical precision, due to the lower signal-to-background ratio than in the decay $\eta\to\pi^+\pi^-\pi^0$. 
We measure the branching fraction of these transitions, and also the ratio of branching fractions:
    \begin{equation}
    {\cal R} = \frac{{\cal B}(\Upsilon(4S)\to\eta\Upsilon(1S))}{{\cal B}(\Upsilon(4S)\to\pi^+\pi^-\Upsilon(1S))}. \label{eq:ratio}
    \end{equation}

The analysis is also potentially sensitive to the transition $\Upsilon(1^3D_{1,2})\to\eta\Upsilon(1S)$, which could be observable in the same final state reconstructed for the $\Upsilon(4S)\to\eta\Upsilon(1S)$ study, with the subsequent decays $\eta\to\pi^+\pi^-\pi^0$, $\pi^0\to\gamma\gamma$, and $\Upsilon(1S)\to\mu^+\mu^-$. The $\Upsilon(1^3D_{1,2})$ could be produced through double-radiative transitions from the $\Upsilon(4S)$ through the $\chi_{bJ}(2P)$ states, while the contribution from the $\Upsilon(3S)$ produced in initial state radiation (ISR) is expected to be negligible. The decay $\Upsilon(1^3D_{1,2})\to\eta\Upsilon(1S)$ has been predicted to be enhanced with respect to the transition $\Upsilon(1^3D_{1,2})\to\pi^+\pi^-\Upsilon(1S)$ by the axial anomaly in QCD~\cite{ref:Voloshin1D}.

\section{Data samples and detector}\label{sec:data}
We use a sample of $(538\pm7)\times10^6$ $\Upsilon(4S)$ mesons, corresponding to the number of $B\bar B$ pairs produced in a sample of integrated luminosity of ${\cal L}_{\rm{int}}=496$ fb$^{-1}$, collected by the Belle experiment at a center-of-mass (CM) energy corresponding to the mass of the $\Upsilon(4S)$ meson at the KEKB asymmetric-energy $e^+e^-$ collider~\cite{ref:KEKB,ref:KEKB_bis}. 
In addition, a data sample corresponding to 56 fb$^{-1}$, collected about 60~MeV below the resonance, is used to estimate the background contribution. Decays of $\Upsilon(3S)$ and $\Upsilon(2S)$ mesons are studied in events recorded at the energy of the $\Upsilon(4S)$ and assumed to come from ISR production; the ISR photon is typically emitted  almost collinear to the beam direction and is not required to be reconstructed. The equivalent luminosity for a narrow vector resonance produced in ISR events is calculated as in Ref.~\cite{ref:ISRlumi}, and is $\sim17.1$ pb and $\sim28.6$ pb for the $\Upsilon(2S)$ and the $\Upsilon(3S)$, respectively.

The Belle detector (described in detail elsewhere~\cite{ref:Belle,ref:Belle_bis})  is a large-solid-angle magnetic spectrometer that consists of a silicon vertex detector, a 50-layer central drift chamber (CDC), an array of aerogel threshold Cherenkov counters (ACC), a barrel-like arrangement of time-of-flight scintillation counters, and an electromagnetic calorimeter comprised of CsI(Tl) crystals (ECL) located inside a super-conducting solenoid coil that provides a 1.5~T magnetic field.  An iron flux-return located outside of the coil is instrumented to detect $K_L^0$ mesons and to identify muons (KLM). 

Monte Carlo (MC) simulated events are used for the efficiency determination and the selection optimization, and are generated with \texttt{EvtGen}~\cite{ref:EvtGen}, while \texttt{GEANT}3~\cite{ref:GEANT} is used to simulate the detector response. The changing detector performance and accelerator conditions are taken into account in the simulation.
The distributions of generated dimuon decays incorporate the $\Upsilon(nS)$ polarization. Dipion transitions as well as $\Upsilon(1^3D_{1,2})\to\eta\Upsilon(1S)$ decays are generated according to phase space, while the angular distribution in $\Upsilon(4S)\to\eta\Upsilon(1S)$ events is simulated as a vector decaying to a pseudoscalar and a vector. The $\eta\to\pi^+\pi^-\pi^0$ decays are modeled according to the known Dalitz plot parameters~\cite{ref:PDG2016}. Final state radiation effects are described by \texttt{PHOTOS}~\cite{ref:PHOTOS}, and secondary emission is taken into account in the simulation of $\Upsilon(3S,2S)$ resonances produced in ISR. 

\section{Event selection}\label{sec:sel}
Charged tracks must originate from a cylindrical region of radius 1 cm and axial length $\pm$5 cm centered on the $e^+e^-$ interaction point and have a momentum transverse to the beam axis ($p_{\rm T}$) greater than 0.1 GeV/$c$, with the $z$ axis chosen to be antiparallel to the $e^+$ beam. Charged particles are assigned a likelihood ${\cal L}_i$ ($i = \mu, \pi, K$)~\cite{ref:muID} based on the range of the particle in the KLM, and on matching it to the track extrapolated from the CDC; particles are identified as muons if the likelihood ratio ${\cal P}_{\mu} = {\cal L}_\mu/({\cal L}_\mu + {\cal L}_\pi + {\cal L}_K)$ exceeds 0.8, corresponding to a muon efficiency of about 91.5$\%$ over the polar angle range $20\degree \leq \theta \leq 155\degree$ and the momentum range 0.7 GeV/$c \leq p \leq 3.0$ GeV/$c$ in the laboratory frame. Electron identification uses a similar likelihood ratio ${\cal P}_e$ based on CDC, ACC, and ECL information~\cite{ref:eID}. Charged particles that are not identified as muons and have a likelihood ratio ${\cal P}_e<0.1$ are treated as pions, thus rejecting $\sim75\%$ of the background events due to photon conversions in the detector material, while retaining almost $99\%$ of the signal. Calorimeter clusters not associated with reconstructed charged tracks and with energies greater than 50 MeV are classified as photon candidates.

Each muon candidate is required to have a CM momentum, $p(\mu)_{\rm CM}$, between 4.25 (4.9) GeV/$c$ and 5.25 (5.1) GeV/$c$ in the case of decays to $\Upsilon(1S)$ ($\Upsilon(2S)$). At least one of the muon candidates must be positively identified as a muon. Pairs of oppositely charged tracks classified as pions are selected to form dipion candidates.
Candidate events must contain a pair of oppositely charged pions, and two muons from the decay of the $\Upsilon(nS)$, the pair having an invariant mass $M({\mu\mu})$  within $\pm4\sigma$ of the known value~\cite{ref:PDG2016} for the considered resonance. This results in requiring  events corresponding to the transitions to $\Upsilon(1S)$ to have 9.2 GeV/$c^2 < M(\mu\mu) < 9.7$ GeV/$c^2$, and events corresponding to the $\Upsilon(4S)\to\pi^+\pi^-\Upsilon(2S)$ transition to have 9.8 GeV/$c^2 < M(\mu\mu) < 10.2$ GeV/$c^2$. 

The quantity $p_{\rm KB} = p(\mu\mu)_{CM} - (s-M(\mu\mu)^2c^4)/(2c\sqrt{s})$, where $p(\mu\mu)_{CM}$ is the CM momentum of the dimuon system and $\sqrt{s}$ is the CM $e^+e^-$ energy, represents a kinematic bound and is expected to be kinematically constrained to negative values for the signal events, and is used to reject most of the background contribution due to QED processes ($e^+e^-\to e^+e^-(\gamma)$ and $e^+e^-\to \mu^+\mu^-(\gamma)$). In the case of dipion transitions, remaining backgrounds are due to QED processes, where a photon converts in the detector material and the leptons are reconstructed as pions. This contribution to the background is reduced by requiring  the opening angle of the charged pion candidates in the laboratory frame to have $\cos\theta(\pi\pi)<0.9$; in addition, the invariant mass $m_{\rm conv}$ of the charged tracks associated with the pion candidates, calculated assuming the $e^\pm$ mass hypothesis, must be greater than 100 MeV/$c^2$. Cosmic background events are typically back-to-back and are rejected by requiring that $\cos\theta(\pi\pi)>-0.98$. 

When looking for $\Upsilon(4S)\to\eta\Upsilon(1S)$ and  $\Upsilon(1^3D_{1,2})\to\eta\Upsilon(1S)$ transitions, only events with at least two additional photons of energy $E_{\gamma}>50$ MeV, invariant mass 110  MeV/$c^2 < M(\gamma\gamma) < 150$ MeV/$c^2$, and with an invariant mass, when combined with the two charged pion candidates, within 50 MeV/$c^2$ of the nominal $\eta$ mass, are retained. The chosen mass windows correspond to $\pm2.5\sigma$ around the nominal $m_{\pi^0}$ and $m_\eta$. The opening angle of the charged pion candidates from the $\eta$ decay in the laboratory frame is required to have $\cos\theta(\pi\pi)>0.5$. An additional requirement $m_{\rm conv}<300$ MeV/$c^2$ helps in reducing the cross-feed from the higher-statistics dipion transitions. Similarly, events with $\Delta M = M(\pi\pi\mu\mu) - M(\mu\mu)$ within 20 MeV/$c^2$ from the values expected for any known dipion transition are vetoed. A significant combinatorial background arises from selecting the incorrect photon candidates for  the $\pi^0$ daughters;  when multiple candidates are present, the ambiguity is resolved by choosing the one whose pair of photons has an invariant mass closest to the nominal $\pi^0$ mass, and that, combined with the two pion candidates, gives an invariant mass closest to the $\eta$ mass.

The criteria applied in the event selection are summarized in Table~\ref{tab:selection}. Table~\ref{tab:efficiency} reports the selection efficiency for all the studied transitions, as determined from MC-simulated samples.

\begin{table*}[htb]
\caption{Summary of event selection criteria.}
\label{tab:selection}
\begin{tabular}
 {@{\hspace{0.5cm}}c@{\hspace{0.5cm}}  @{\hspace{0.5cm}}c@{\hspace{0.5cm}}}
\hline \hline
$\Upsilon(4S)\to\pi^+\pi^-\Upsilon(2S)$ & Other dipion transitions\\
\hline
$p_{\rm KB} < 0$ GeV/$c$ & $p_{\rm KB}<-0.1$ GeV/$c$\\
4.9 GeV/$c < p(\mu)_{\rm CM} < 5.1$ GeV/$c$ & 4.25 GeV/$c < p(\mu)_{\rm CM} < 5.25$ GeV/$c$\\
$-0.98< \cos\theta(\pi\pi) <0.9$ & $-0.98< \cos\theta(\pi\pi) <0.9$\\
$m_{\rm conv}<500$ MeV/$c^2$ & $m_{\rm conv}>100$ MeV/$c^2$\\
\hline \hline
 $\Upsilon(4S)\to\eta\Upsilon(1S)$ & $\Upsilon(1^3D_{1,2})\to\eta\Upsilon(1S)$\\
\hline
$p_{\rm KB}<-0.1$ GeV/$c$ & $p_{\rm KB}<-0.3$ GeV/$c$\\
4.25 GeV/$c < p(\mu)_{\rm CM} < 5.25$ GeV/$c$ & 4.25 GeV/$c < p(\mu)_{\rm CM} < 5.25$ GeV/$c$\\
$\cos\theta(\pi\pi) >0.5$ & 0.5 $< \cos\theta(\pi\pi) < 0.9$\\
100 MeV/$c^2$ $< m_{\rm conv}<300$ MeV/$c^2$ & 100 MeV/$c^2$ $< m_{\rm conv}<300$ MeV/$c^2$\\
\hline \hline 
\end{tabular}
\end{table*}

\begin{table}[htb]
\caption{Selection efficiency ($\epsilon$) values for all the studied transitions, as determined from MC-simulated samples. For the dipion transitions, the phase-space averaged efficiency is reported. The $\Upsilon(1^3D_{1,2})$ is intended to be produced in $\Upsilon(4S)\to\gamma\gamma\Upsilon(1^3D_{1,2})$ events.}
\label{tab:efficiency}
\begin{tabular}
 {@{\hspace{0.5cm}}l@{\hspace{0.5cm}}  @{\hspace{0.5cm}}r@{\hspace{0.5cm}}}
\hline \hline
Transition & Selection efficiency ($\%$)\\
\hline
$\Upsilon(2S)\to\pi^+\pi^-\Upsilon(1S)$ & $29.63\pm 0.05$\\
$\Upsilon(3S)\to\pi^+\pi^-\Upsilon(1S)$ & $43.52\pm 0.05$\\
$\Upsilon(4S)\to\pi^+\pi^-\Upsilon(1S)$ & $47.49\pm 0.05$\\
$\Upsilon(4S)\to\pi^+\pi^-\Upsilon(2S)$ & $18.27\pm 0.05$\\
\hline
$\Upsilon(4S)\to\eta\Upsilon(1S)$ & $11.46\pm 0.11$\\
$\Upsilon(1^3D_{1,2})\to\eta\Upsilon(1S)$ & $5.72\pm0.08$\\
\hline \hline
\end{tabular}
\end{table}

\section{Signal extraction}\label{sec:fit}
\begin{figure}[htb]
\includegraphics[width=0.44\textwidth]{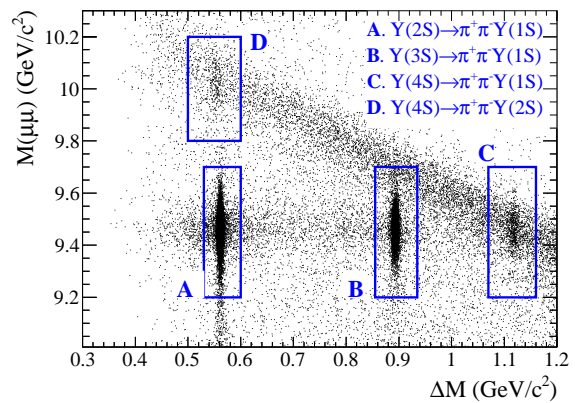}
\caption{Distribution of $M(\mu\mu)$ vs.\ $\Delta M$ for the events selected on data. Fit regions for the four analyzed dipion transitions are enclosed in boxes.}\label{fig:2D_dip}
\end{figure}

For the dipion transitions, the two-dimensional distribution of the invariant dimuon mass $M(\mu\mu)$ vs.\ $\Delta M$ for the selected data events is shown in Fig.~\ref{fig:2D_dip}, with the four different decays of interest highlighted. The signal yields are extracted in the four regions shown.

In order not to introduce any bias in the assumptions on the angular distribution of the decay, the signal yield is separately estimated and corrected for the efficiency in $6\times4$ bins of $M(\pi^+\pi^-)$ and $\cos\theta_{\rm hel}(\pi^+)$ for the $\Upsilon(2S,3S)\to\pi^+\pi^-\Upsilon(1S)$ transitions,  where $M(\pi^+\pi^-)$ is the invariant mass of the dipion system and $\theta_{\rm hel}(\pi^+)$ represents the helicity angle of the positive pion candidate, defined as the angle between the $\pi^+$ direction and the recoiling lower-mass $\Upsilon$ in the dipion rest frame. For the lower-statistics $\Upsilon(4S)\to\pi^+\pi^-\Upsilon(2S,1S)$ transitions, $4\times4$ bins are used. 
In each bin, the signal and background yields are determined by an unbinned maximum likelihood fit to the $\Delta M$ distribution. The signal component is parameterized by a Voigtian function, with the resolution parameters fixed to the values determined from the MC-simulated samples. The background is parameterized by a linear function.

\begin{figure*}[htb]
\includegraphics[width=0.44\textwidth]{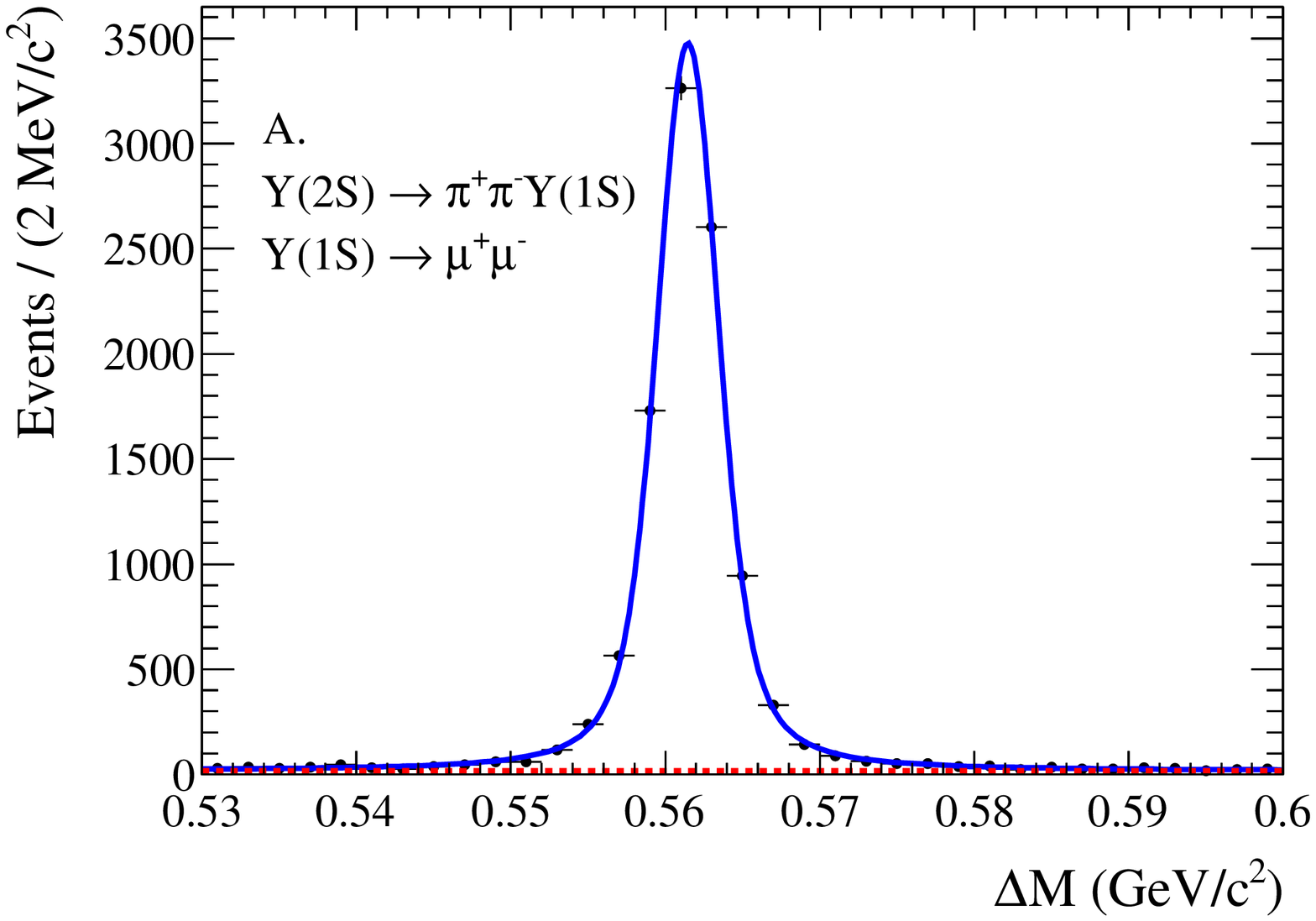}\includegraphics[width=0.44\textwidth]{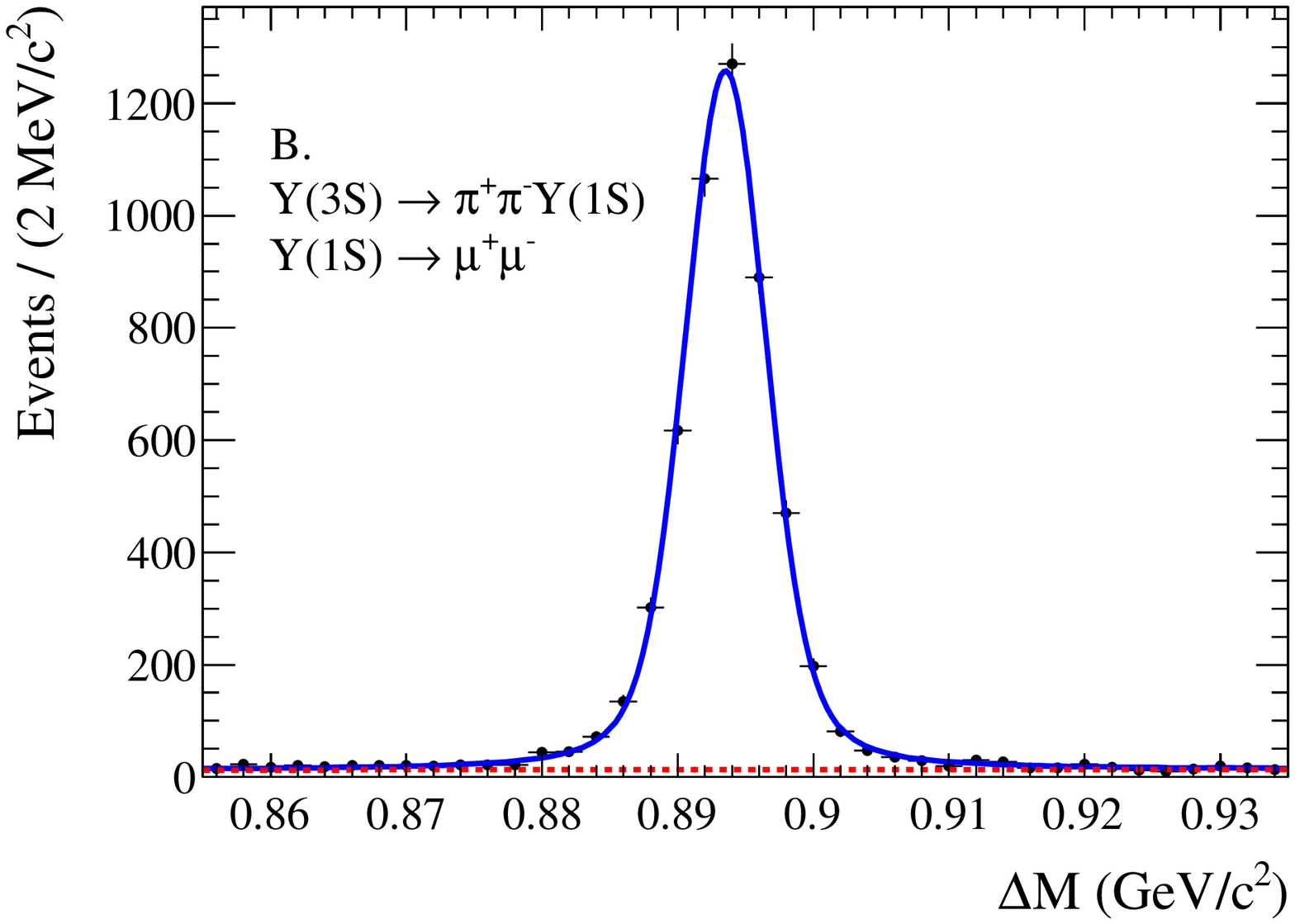}
\includegraphics[width=0.44\textwidth]{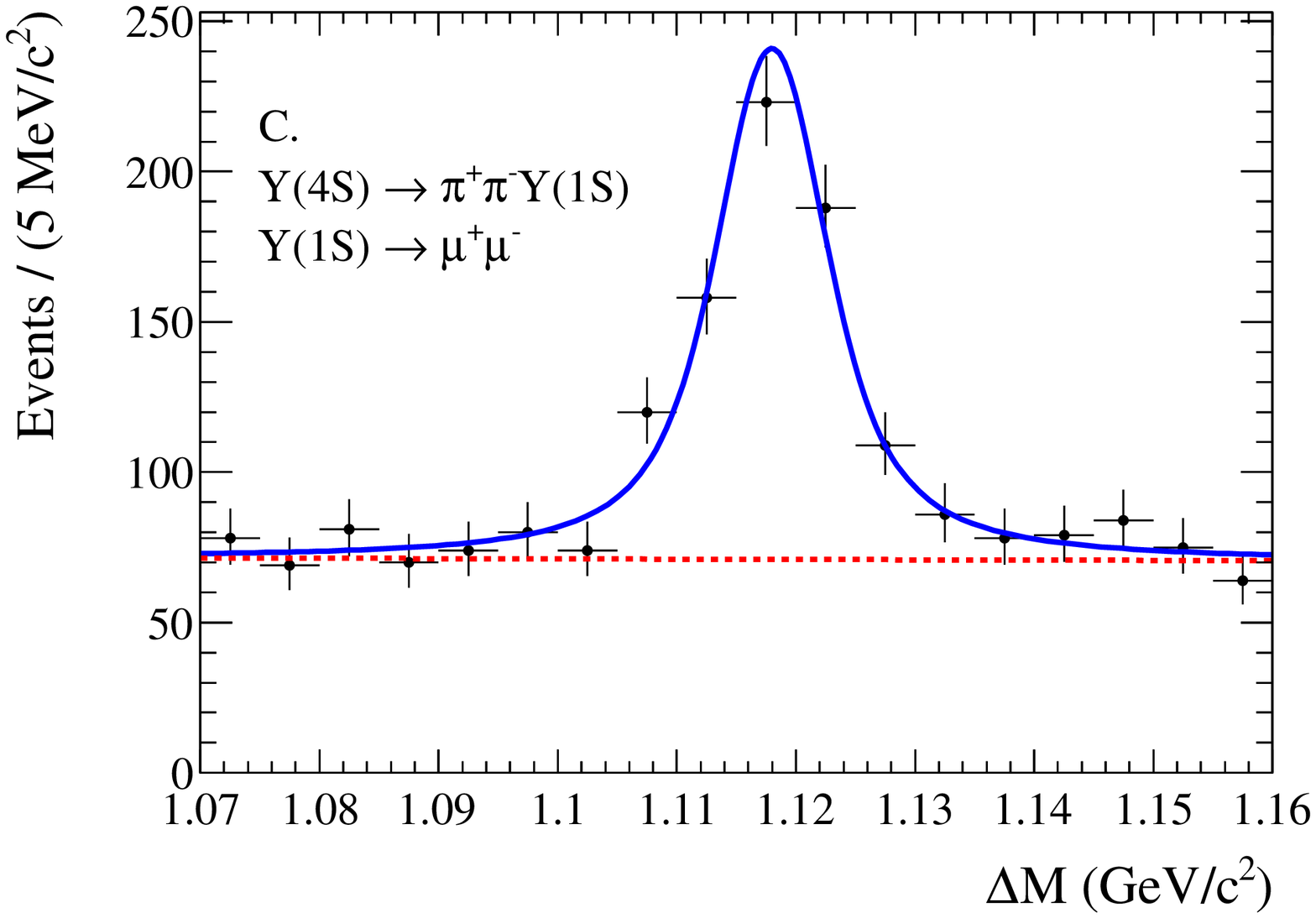}\includegraphics[width=0.44\textwidth]{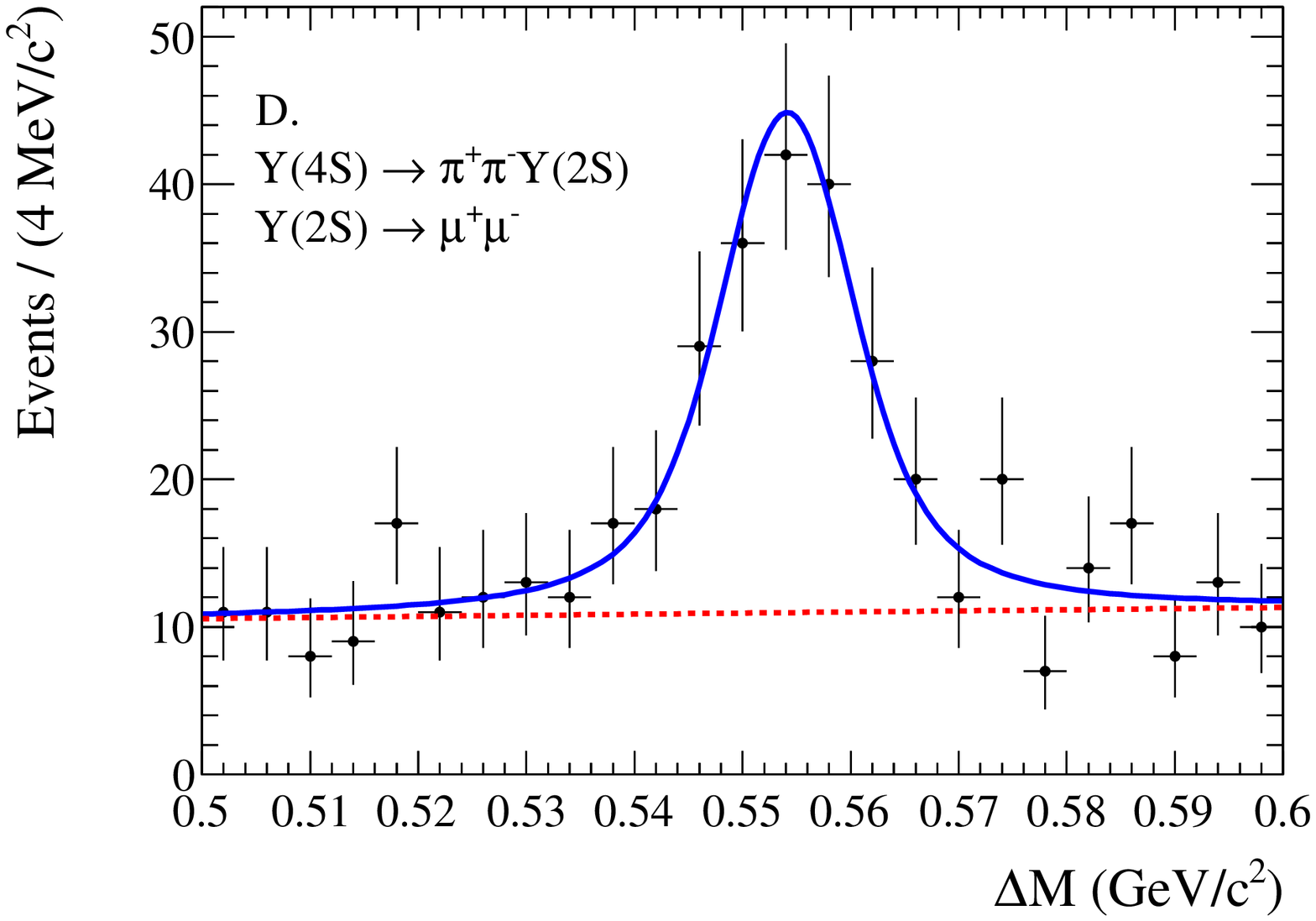}
\caption{Fits to the $\Delta M$ distributions for $\Upsilon(2S,3S,4S)\to\pi^+\pi^-\Upsilon(1S,2S)$ candidates. In each plot, data are shown as points, the solid blue line shows the best fit to the data, while the dashed red line shows the background contribution.}\label{fig:fit_dip}
\end{figure*}

For each transition, the efficiency-corrected signal yield is estimated as $N_{\rm corrected} = \sum_{\rm bins} N_{\rm sig}^i/\epsilon_i$ where the sum is over all of the considered bins, and $N_{\rm sig}^i$ and $\epsilon_i$ are, respectively, the signal yield, determined from the fit, and the efficiency, obtained from MC samples, in the $i^{\rm th}$ bin. The results are listed in Table~\ref{tab:yields}, and the distributions of $\Delta M$ for the selected data events, integrated over the $M(\pi^+\pi^-)$ vs.\ $\cos\theta_{\rm hel}(\pi^+)$ bins, are shown in Fig.~\ref{fig:fit_dip}.

\begin{figure}[htb]
\includegraphics[width=0.44\textwidth]{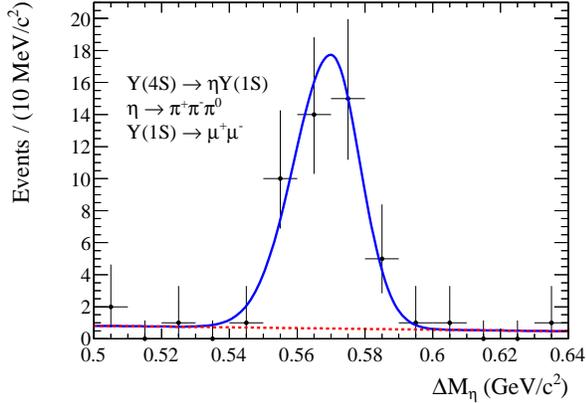}
\caption{Fit to the $\Delta M_\eta$ distribution for $\Upsilon(4S)\to\eta\Upsilon(1S)$ candidates. Data are shown as points, the solid blue line shows the best fit to the data, while the dashed red line shows the background contribution.}\label{fig:fit_eta}
\end{figure}

\begin{figure}[htb]
\includegraphics[width=0.44\textwidth]{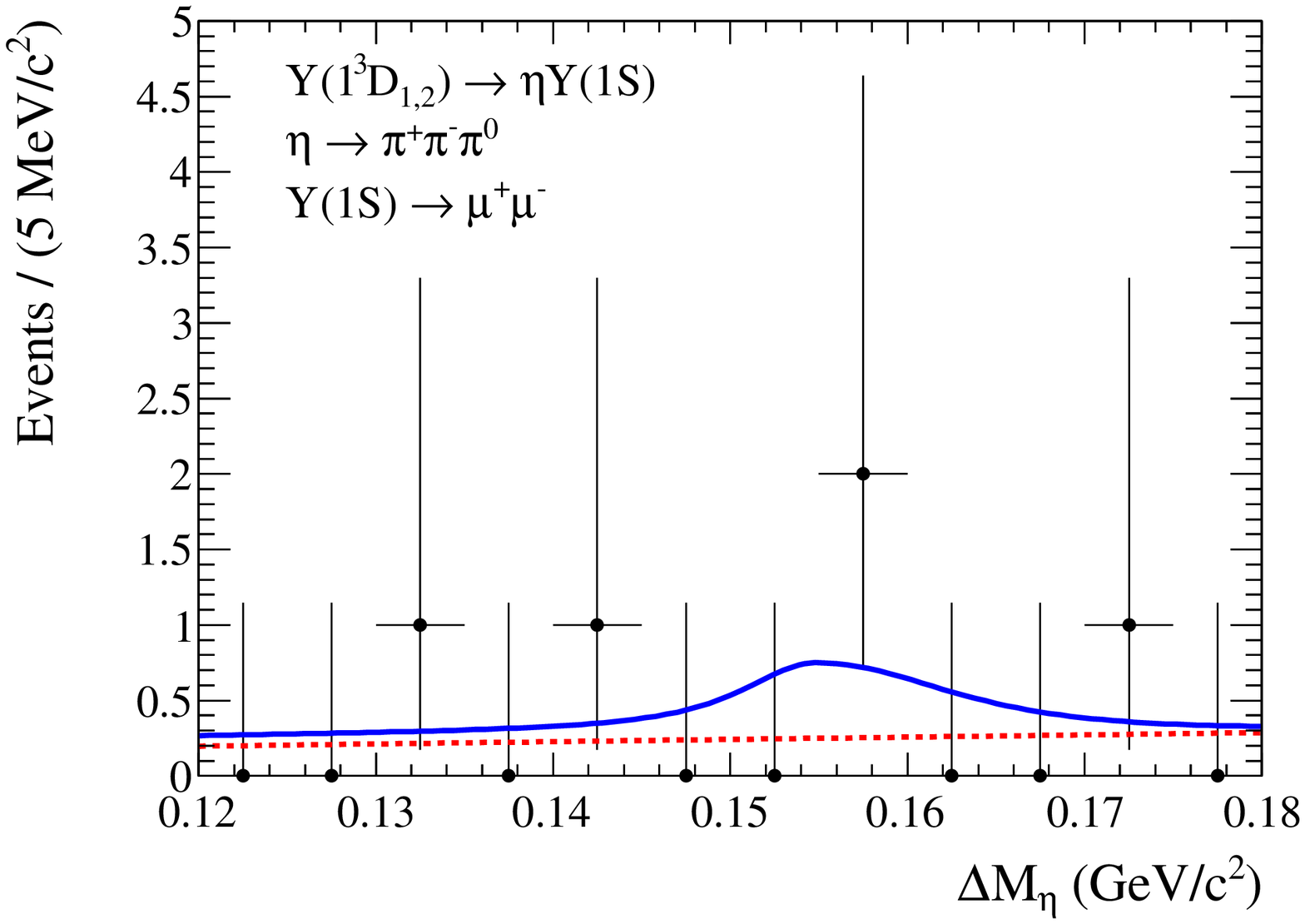}
\caption{Fit to the $\Delta M_\eta$ distribution for $\Upsilon(1^3D_{1,2})\to\eta\Upsilon(1S)$ candidates. Data are shown as points, the solid blue line shows the best fit to the data, while the dashed red line shows the background contribution.}\label{fig:fit_eta1D}
\end{figure}

For the $\Upsilon(4S)\to\eta\Upsilon(1S)$ transition, the distribution of $\Delta M_\eta =  M(\pi\pi\gamma\gamma\mu\mu) - M(\mu\mu) - M(\pi\pi\gamma\gamma)$ for the selected data events is shown in Fig.~\ref{fig:fit_eta}, with 51 candidate events  found in the fit region 0.50 GeV/$c^2<\Delta M_\eta <$ 0.64 GeV/$c^2$. For the $\Upsilon(1^3D_{1,2})\to\eta\Upsilon(1S)$ transition, the distribution of $\Delta M_\eta$ for the selected data events is shown in Fig.~\ref{fig:fit_eta1D}, with 5 candidate events  found in the fit region 0.12 GeV/$c^2<\Delta M_\eta <$ 0.18 GeV/$c^2$. 
The signal and background yields are determined by an unbinned maximum likelihood fit to this distribution. For both transitions, the signal component is parameterized by a Gaussian-like analytical function, with mean value $\mu$ and different widths, $\sigma_{\text{L,R}}$, on the left side (for $x<\mu$) and on the right side (for $x>\mu$) plus asymmetric tails $\alpha_{\text{L,R}}$, defined as:
    \begin{equation}
    {\cal F}(x) = \exp\Big\{-\frac{(x-\mu)^2}{2\sigma_{\text{L,R}}^2+\alpha_{\text{L,R}}(x-\mu)^2}\Big\}. \label{eq:cruijffpdf}
    \end{equation}
The background is described by a linear function. For the $\Upsilon(4S)\to\eta\Upsilon(1S)$ transition, all the parameters of the functional forms describing the signal and the background components are left free to vary in the fit, while, for  the $\Upsilon(1^3D_{1,2})\to\eta\Upsilon(1S)$ transition, the signal shape parameters are fixed to the values determined on the MC simulated sample. The signal and background yields are reported in Table~\ref{tab:yields}.

\begin{table}[htb]
\caption{Signal and background yields for the analyzed transitions. $N_{\rm bkg}$ is the number of background events, in the entire fit region. For the transition with an $\eta$ meson, $N_{\rm sig}$ is the number of signal events in the entire fit region. For the dipion transitions, $N_{\rm sig}=\sum_{\rm bins} N_{\rm sig}^i$ is the sum of the signal yields obtained in each bin ($i^{\rm th}$), without corrections for the efficiency; the efficiency-corrected yields are shown as $N_{\rm corrected}$, as defined in Sec.~\ref{sec:fit}.}
\label{tab:yields}
\begin{tabular}
 {@{\hspace{0.05cm}}l@{\hspace{0.05cm}} @{\hspace{0.05cm}}|r@{\hspace{0.1cm}}@{\hspace{0.01cm}}c@{\hspace{0.01cm}}@{\hspace{0.1cm}}r@{\hspace{0.1cm}}@{\hspace{0.1cm}}r@{\hspace{0.1cm}}@{\hspace{0.01cm}}c@{\hspace{0.01cm}}@{\hspace{0.1cm}}r@{\hspace{0.1cm}} @{\hspace{0.1cm}}r@{\hspace{0.1cm}}@{\hspace{0.01cm}}c@{\hspace{0.01cm}}@{\hspace{0.1cm}}r@{\hspace{0.05cm}}}
\hline \hline
Transition & \multicolumn{3}{c}{$N_{\rm sig}$} & \multicolumn{3}{c}{$N_{\rm corrected}$} & \multicolumn{3}{c}{$N_{\rm bkg}$}\\
\hline
$\Upsilon(2S)\to\pi^+\pi^-\Upsilon(1S)$ & 9805&$\pm$&106 & 38117&$\pm$&419 & 287&$\pm$&41\\
$\Upsilon(3S)\to\pi^+\pi^-\Upsilon(1S)$ & 5222&$\pm$&77 & 15526&$\pm$&252 & 518&$\pm$&33\\
$\Upsilon(4S)\to\pi^+\pi^-\Upsilon(1S)$ & 515&$\pm$&34 & 1095&$\pm$&74 & 1278&$\pm$&45\\
$\Upsilon(4S)\to\pi^+\pi^-\Upsilon(2S)$ & 181&$\pm$&20 & 821&$\pm$&107 & 273&$\pm$&22\\
\hline
$\Upsilon(4S)\to\eta\Upsilon(1S)$ & 49&$\pm$&7  & \multicolumn{3}{c}{ } & 2.3&$\pm$&1.8\\
$\Upsilon(1^3D_{1,2})\to\eta\Upsilon(1S)$ & 2.1&$\pm$&3.0  & \multicolumn{3}{c}{ } & 2.9&$\pm$&3.1\\
\hline \hline
\end{tabular}
\end{table}

\section{Systematic uncertainties}\label{sec:syst}
The sources of systematic uncertainty affecting our measurement are itemized here. An uncertainty comes from the number of $\Upsilon(4S)$ parents and from the values used for the secondary branching fractions~\cite{ref:PDG2016}. The uncertainties in charged track reconstruction and muon identification efficiency are determined by comparing data and MC events using independent control samples. 
Another contribution to the uncertainty accounts for the systematic discrepancy between data and MC in the $\pi^0$ reconstruction efficiency.

One of the largest contributions to the systematic uncertainty comes from the signal extraction procedure. The uncertainty due to the choice of signal parameterizations is estimated by changing the functional forms used; the systematic uncertainty on the background description is evaluated by using higher-order polynomial functions while enlarging the range chosen for the fit. For the dipion transitions, additional sources of systematic uncertainties have been taken into account. A systematic discrepancy in the resolution between data and MC is evaluated by floating independently the resolution parameters of the functional form describing the signal. Finally, the uncertainty in the acceptance correction is determined by using different numbers of bins in $M(\pi^+\pi^-)$ and $\cos\theta_{\rm hel}(\pi^+)$. In each case, the uncertainty is estimated as the change in the signal yield when using an alternate configuration with respect to that obtained with the nominal one.

Other possible sources of systematic uncertainties associated wth the event selection and due to discrepancies between data and MC in the efficiency of the applied requirements, have been found to be negligible.

All the considered sources of systematic uncertainty are summarized in Table~\ref{tab:syst}, for each transition. The total systematic uncertainty is obtained by adding in quadrature all the contributions. When measuring the ratio given in Eq.~\ref{eq:ratio}, several systematic uncertainties cancel, being common to the numerator and the denominator of the ratio; these contributions are specifically indicated in Table~\ref{tab:syst} and sum up to $5.3\%$.

\begin{table*}[htb]
\caption{Systematic uncertainties on branching fractions, in percent. 
The sources contributing to the measurement of the ratio in Eq.~\ref{eq:ratio} are underlined. The $\oplus$ symbol indicates that the two contributions (only one of which contributes to the measurement of the ratio) are added in quadrature.}
\label{tab:syst}
\begin{tabular}
 {@{\hspace{0.3cm}}l@{\hspace{0.3cm}}  @{\hspace{0.3cm}}c@{\hspace{0.3cm}} @{\hspace{0.3cm}}c@{\hspace{0.3cm}} @{\hspace{0.3cm}}c@{\hspace{0.3cm}} @{\hspace{0.3cm}}c@{\hspace{0.3cm}} @{\hspace{0.3cm}}c@{\hspace{0.3cm}}}
\hline \hline
  & $\Upsilon(2S)\to$   & $\Upsilon(3S)\to$  & \multicolumn{3}{c}{$\Upsilon(4S)\to$}\\
Source & $\pi^+\pi^-\Upsilon(1S)$ & $\pi^+\pi^-\Upsilon(1S)$ & $\pi^+\pi^-\Upsilon(1S)$ &  $\pi^+\pi^-\Upsilon(2S)$ & $\eta\Upsilon(1S)$\\
\hline
Number of $\Upsilon(4S)$ & 1.4 & 1.4 & 1.4 & 1.4 & 1.4\\
Secondary BRs            & 2.5 & 2.7 & 2.0 & 2.0 & $2.0\oplus\underline{1.2}$\\
Tracking                             & 1.4 & 1.4 & 1.4 & 1.4 & 1.4\\
$\mu$-identification           & 1.1 & 1.1 & 1.1 & 1.1 & 1.1\\
Signal extraction                & 1.9 & 2.7 & $\underline{2.7}$ & 2.7 & $\underline{2.8}$ \\
Acceptance                        & 1.0 & 1.0 & $\underline{3.1}$ & 3.3 & - \\
$\pi^0$ reconstruction    & - & - & - & - & $\underline{1.4}$\\
\hline
Total            & 4.0 & 4.5 & 5.1 & 5.2 & 4.5 \\
\hline \hline
\end{tabular}
\end{table*}

\section{Results}\label{sec:res}
The results for the branching fractions of $\Upsilon(4S)$ hadronic transitions and the ratio of branching fractions (Eq.~\ref{eq:ratio}) are listed in Table~\ref{tab:results}. They are obtained from the signal yield given in each mode by the fit procedure, as listed in Table~\ref{tab:yields}, eventually efficiency-corrected for the dipion transitions, as explained in Sec.~\ref{sec:fit}. Since the yields in a data sample collected 60 MeV below the resonance have been checked to be consistent with zero, the number of events observed are attributed to the $\Upsilon(4S)$ decay. The number of $\Upsilon(4S)$ parents is also taken into account in the calculation, as well as the secondary branching fractions. 
The measurements show both the statistical and the systematic errors, the latter estimated as explained in Sec.~\ref{sec:syst}.
The results can be also expressed in terms of visible cross sections, given by the efficiency-corrected signal yield divided by the integrated luminosity: $\sigma(e^+e^-\to\pi^+\pi^-\Upsilon(1S)) = (2.20 \pm 0.13 \pm 0.10)$ fb, $\sigma(e^+e^-\to\pi^+\pi^-\Upsilon(2S)) = (1.64 \pm 0.17 \pm 0.08)$ fb, and $\sigma(e^+e^-\to\eta\Upsilon(1S)) = (1.03 \pm 0.14 \pm 0.04)$ fb, where the first errors are statistical and the second systematic.

In Table~\ref{tab:results}, we also give a comparison of our measurements to the previous world averages, as in~\cite{ref:PDG2016}. All the results are found to be compatible with the previous ones, with a slight improvement in the precision with respect to the measurement by BaBar~\cite{ref:BaBar4S} and the previous measurement of ${\cal B}(\Upsilon(4S)\to\pi^+\pi^-\Upsilon(1S))$ by Belle~\cite{ref:Belle4Sdipion}. 
This work confirms the enhancement of the transition from $\Upsilon(4S)$ to $\Upsilon(1S)$ via the spin-flip exchange of an $\eta$ meson with respect to that proceeding through the emission of a dipion.

The world average branching fractions ${\cal B}(\Upsilon(2S,3S)\to\pi^+\pi^-\Upsilon(1S))$~\cite{ref:PDG2016}, whose precision is dominated by measurements obtained with dedicated higher-statistics data samples, are used for determining the ISR production cross sections of the $\Upsilon(2S,3S)$ resonances: $\sigma_{\rm{ISR}}(\Upsilon(2S,3S)) = N_{\rm corrected}/({\cal B}(\Upsilon(2S,3S)\to\pi^+\pi^-\Upsilon(1S))\times{\cal B}(\Upsilon(1S)\to\mu^+\mu^-)\times{\cal L}_{\rm int})$. The results are listed as well in  Table~\ref{tab:results}, and compared with the values calculated as in Ref.~\cite{ref:ISRlumi}. The uncertainty on the expected values is the experimental uncertainty on the $\Upsilon(2S,3S)\to e^+e^-$ partial width~\cite{ref:PDG2016}.

For the transition $\Upsilon(1^3D_{1,2})\to\eta\Upsilon(1S)$, we do not observe any statistically significant signal, and we set an upper limit, using the Feldman-Cousins method~\cite{ref:Feldman}, on the product of branching fractions ${\cal B}(\Upsilon(4S)\to\gamma\gamma\Upsilon(1^3D_{1,2})) \times {\cal B}(\Upsilon(1^3D_{1,2})\to\eta\Upsilon(1S)) < 2.3\times 10^{-5}$, at 90$\%$ confidence level.

\begin{table*}[htb]
\caption{Results for the branching fractions of $\Upsilon(4S)$ hadronic transitions, and for the ratio given in Eq.~\ref{eq:ratio}, in comparison to previous measurements~\cite{ref:PDG2016}, and results for the ISR production cross sections of $\Upsilon(2S,3S)$, in comparison to the values calculated as in Ref.~\cite{ref:ISRlumi}. The first error is statistical, while the second is systematic.}
\label{tab:results}
\begin{tabular}
 {@{\hspace{0.3cm}}l@{\hspace{0.3cm}}  @{\hspace{0.1cm}}r@{\hspace{0.1cm}}@{\hspace{0.01cm}}c@{\hspace{0.01cm}}@{\hspace{0.1cm}}l@{\hspace{0.01cm}}@{\hspace{0.1cm}}c@{\hspace{0.1cm}}@{\hspace{0.01cm}}l@{\hspace{0.01cm}}@{\hspace{0.1cm}}l@{\hspace{0.1cm}}@{\hspace{0.1cm}}r@{\hspace{0.1cm}}@{\hspace{0.01cm}}c@{\hspace{0.01cm}}@{\hspace{0.1cm}}l@{\hspace{0.1cm}}@{\hspace{0.1cm}}l@{\hspace{0.1cm}}}
\hline \hline
Measurement & \multicolumn{6}{c}{Result} & \multicolumn{4}{c}{PDG value~\cite{ref:PDG2016}}\\
\hline
${\cal B}(\Upsilon(4S)\to\pi^+\pi^-\Upsilon(1S))$ & (8.2&$\pm$&0.5&$\pm$&0.4)&$\times10^{-5}$ & (8.1&$\pm$&0.6)&$\times10^{-5}$\\
${\cal B}(\Upsilon(4S)\to\pi^+\pi^-\Upsilon(2S))$ & (7.9&$\pm$&1.0&$\pm$&0.4)&$\times10^{-5}$ & (8.6&$\pm$&1.3)&$\times10^{-5}$\\
\hline
${\cal B}(\Upsilon(4S)\to\eta\Upsilon(1S))$ & (1.70&$\pm$&0.23&$\pm$&0.08)&$\times10^{-4}$ & (1.96&$\pm$&0.28)&$\times10^{-4}$\\
\hline
${\cal R}$ as in Eq.~\ref{eq:ratio} & 2.07&$\pm$&0.30&$\pm$&0.11&  & 2.41&$\pm$&0.42&  \\
\hline \hline
Measurement & \multicolumn{6}{c}{Result} & \multicolumn{4}{c}{Expected value~\cite{ref:ISRlumi}}\\
\hline
$\sigma_{\rm{ISR}}(\Upsilon(2S))$ & (17.36&$\pm$&0.19&$\pm$&0.69)&pb & (17.1&$\pm$&0.3)&pb\\
$\sigma_{\rm{ISR}}(\Upsilon(3S))$ & (28.9&$\pm$&0.5&$\pm$&1.3)&pb & (28.6&$\pm$&0.5)&pb\\
\hline \hline
\end{tabular}
\end{table*}

\begin{figure*}[htb]
\includegraphics[width=0.44\textwidth]{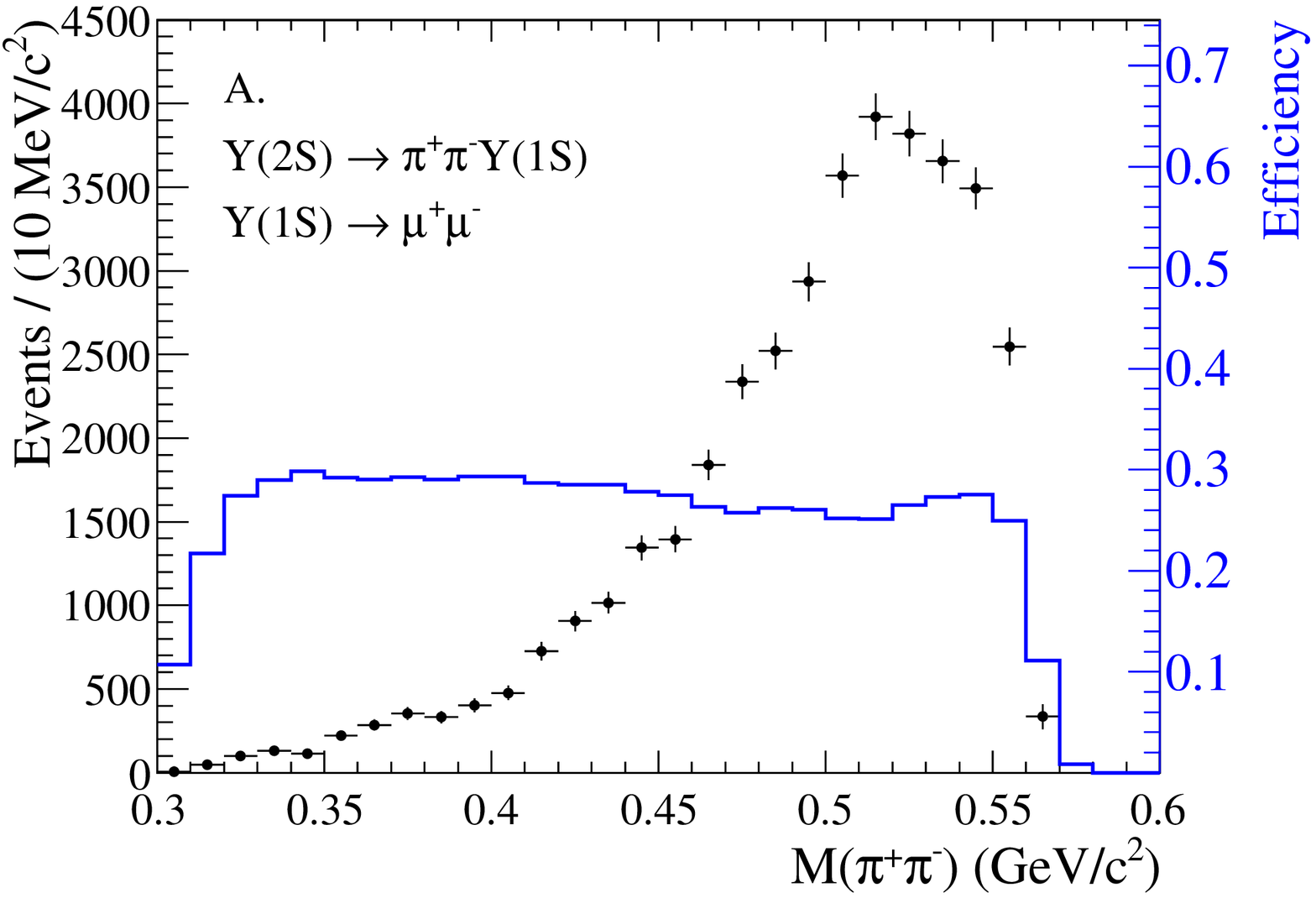}\hspace{1cm}\includegraphics[width=0.44\textwidth]{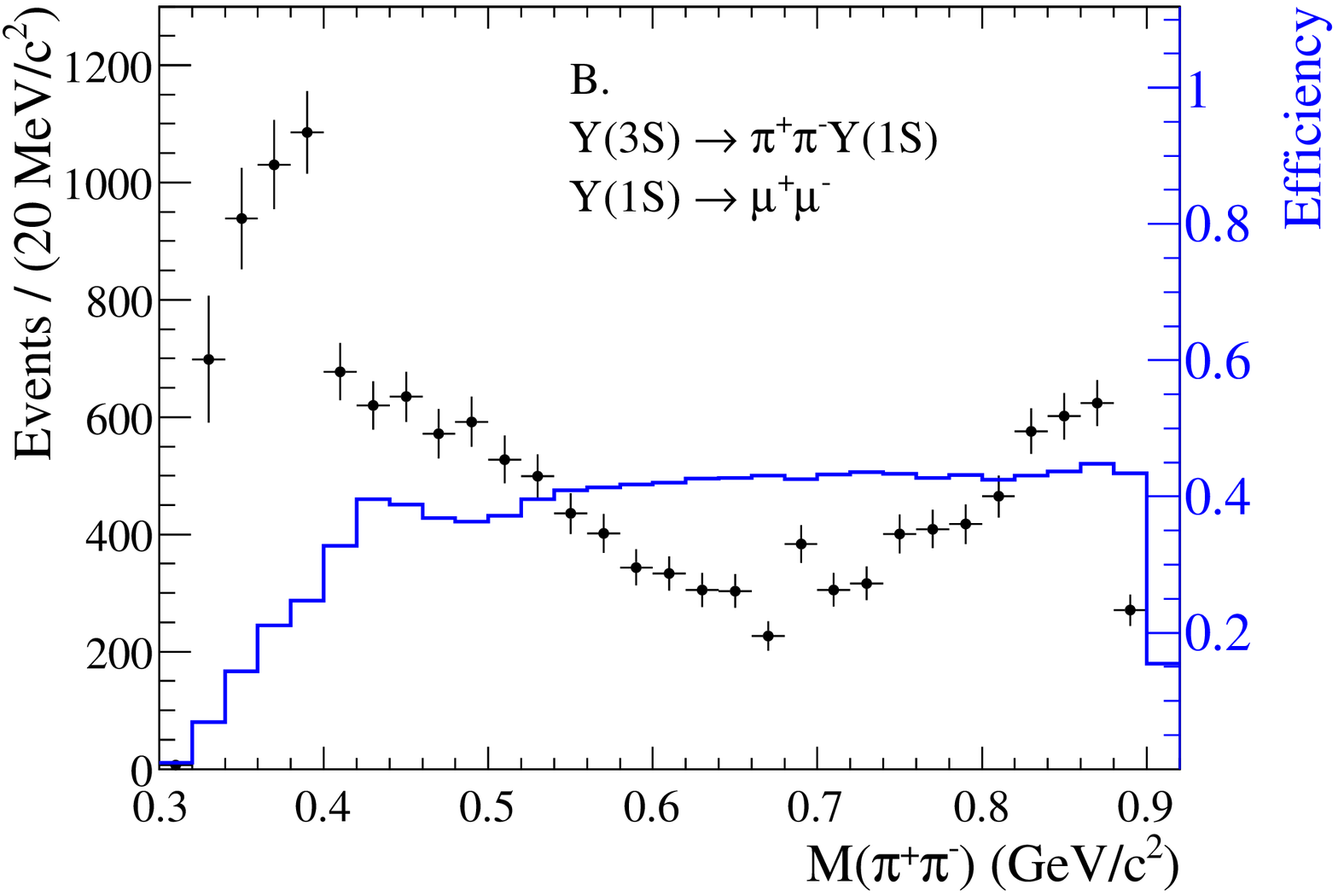}
\includegraphics[width=0.44\textwidth]{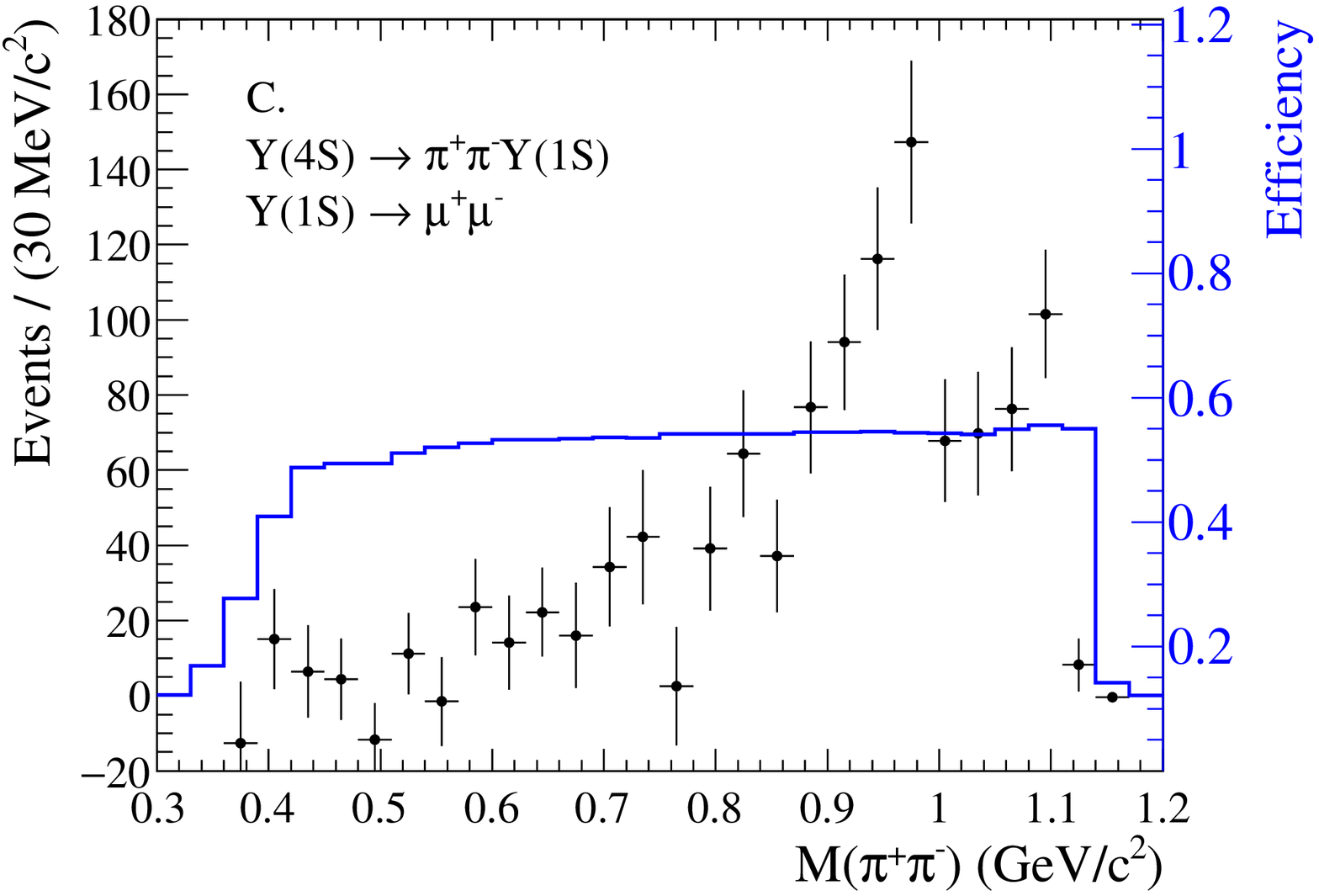}\hspace{1cm}\includegraphics[width=0.44\textwidth]{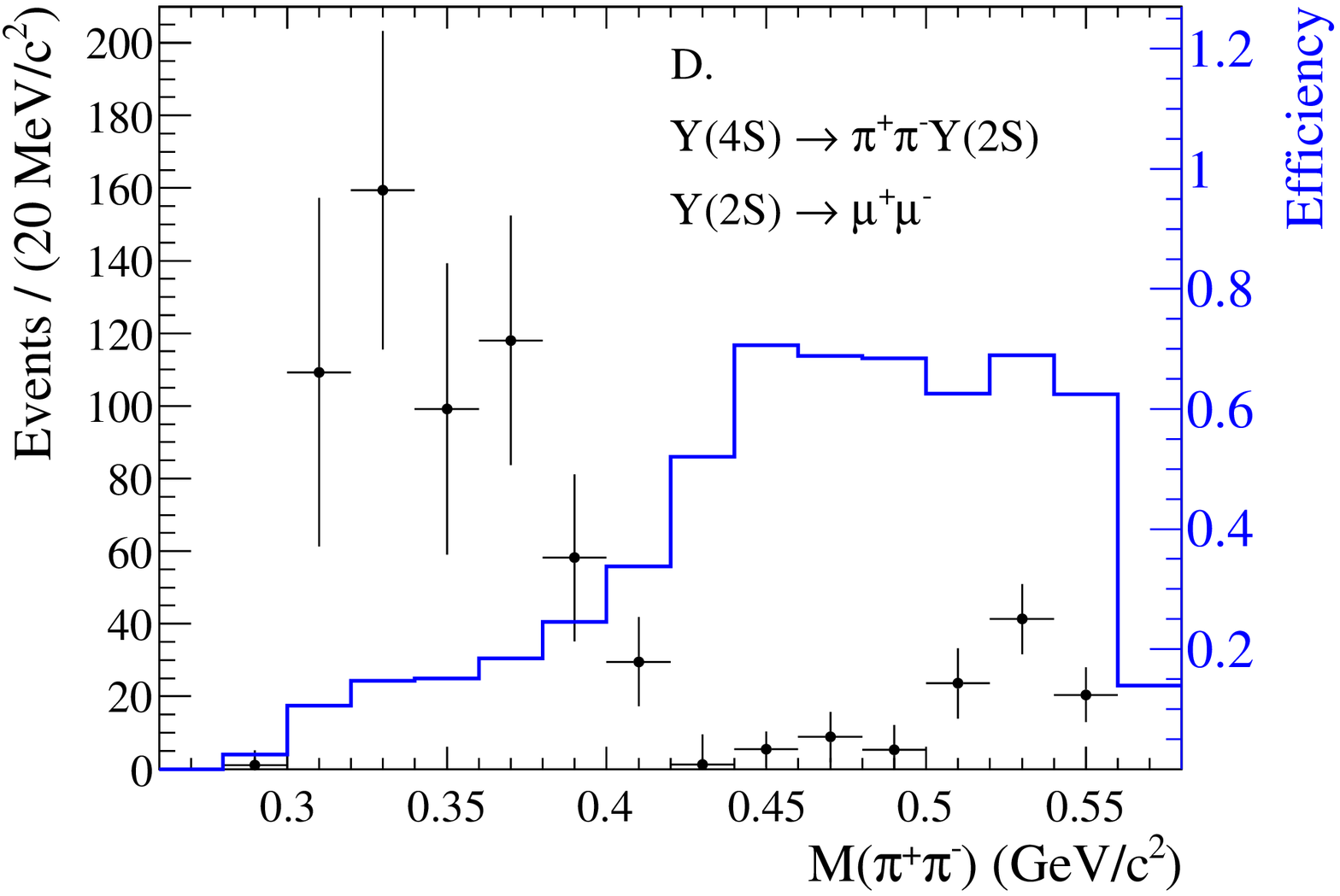}
\caption{Efficiency-corrected distributions of dipion invariant mass ($M(\pi^+\pi^-$)) for the signal component unfolded from the data distributions with the \textit{$_s$}${\cal P}$\textit{lot} technique~\cite{ref:splot} in the $\Upsilon(2S,3S,4S)\to\pi^+\pi^-\Upsilon(1S,2S)$ candidates. The values of the selection efficiency in each bin are shown in the blue histogram (right axis).}\label{fig:splot_mass}
\end{figure*}

\begin{figure*}[htb]
\includegraphics[width=0.44\textwidth]{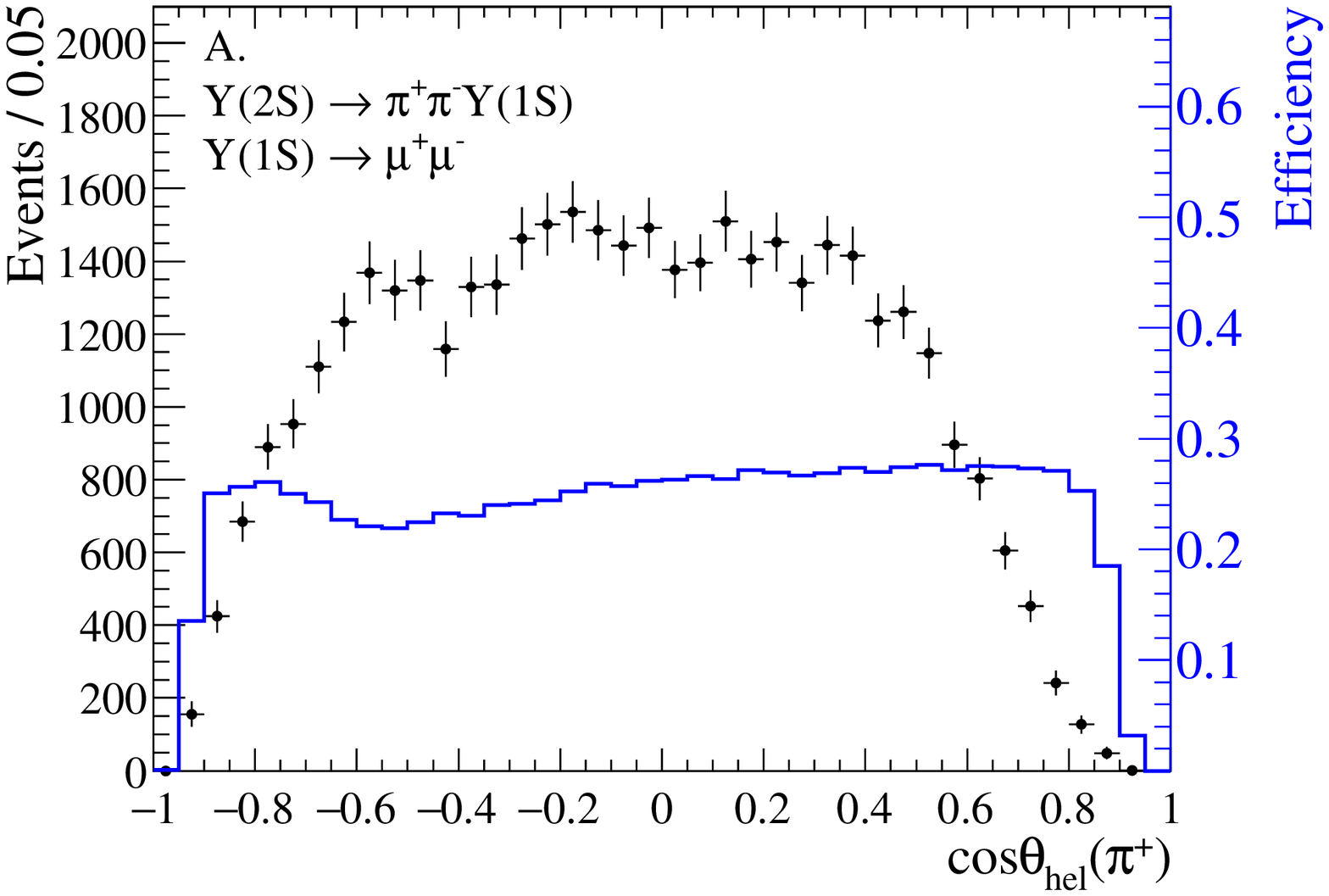}\hspace{1cm}\includegraphics[width=0.44\textwidth]{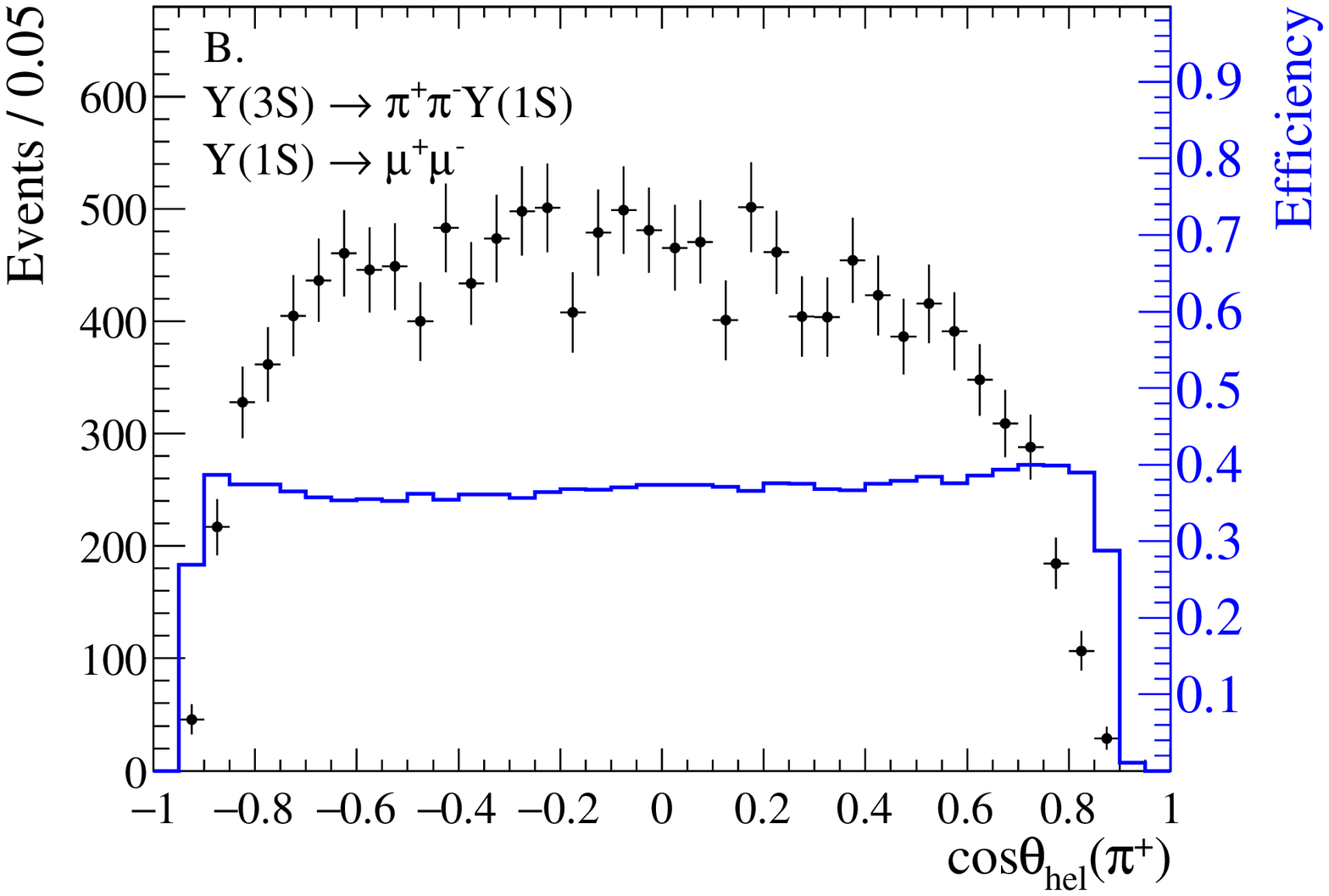}
\includegraphics[width=0.44\textwidth]{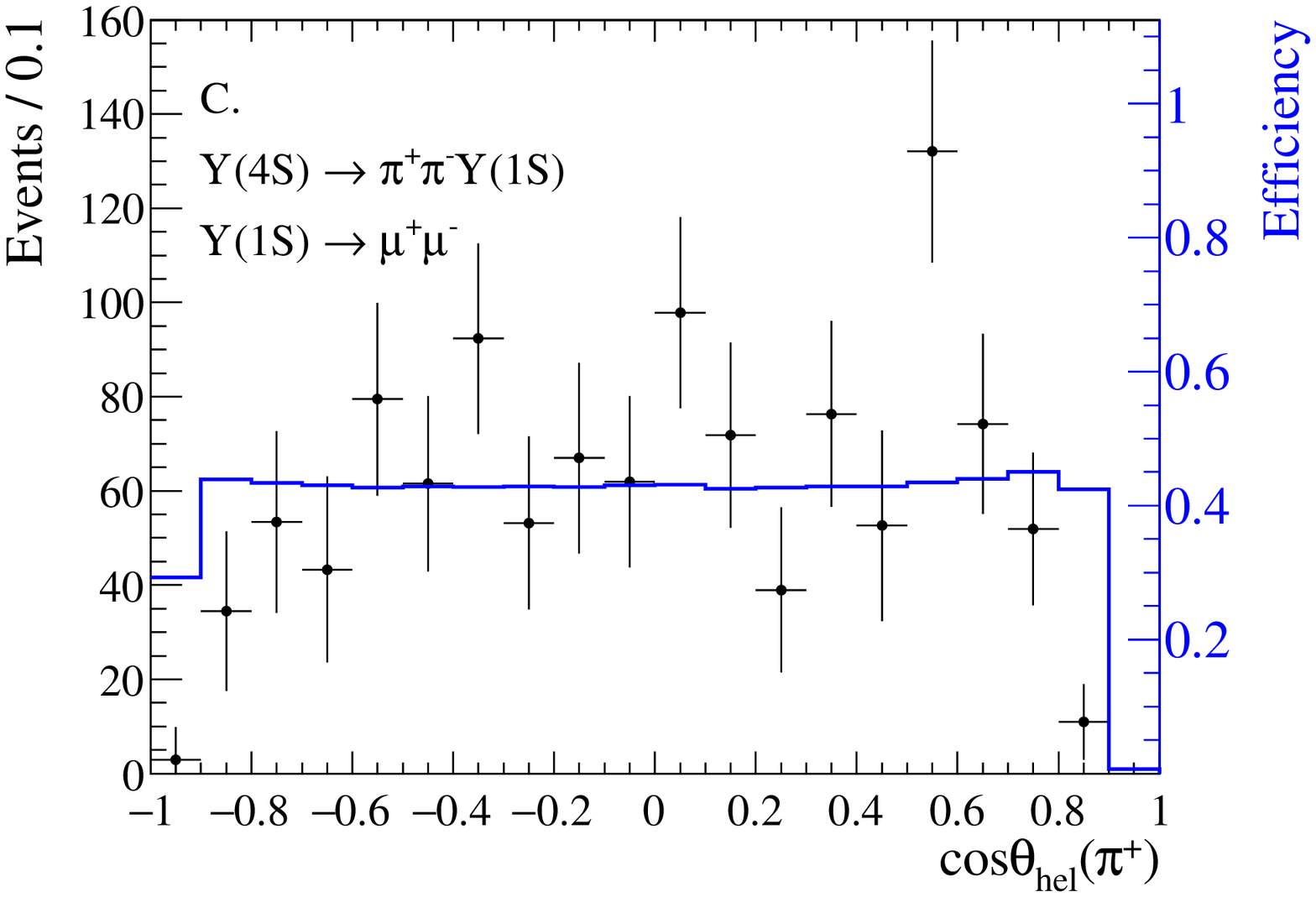}\hspace{1cm}\includegraphics[width=0.44\textwidth]{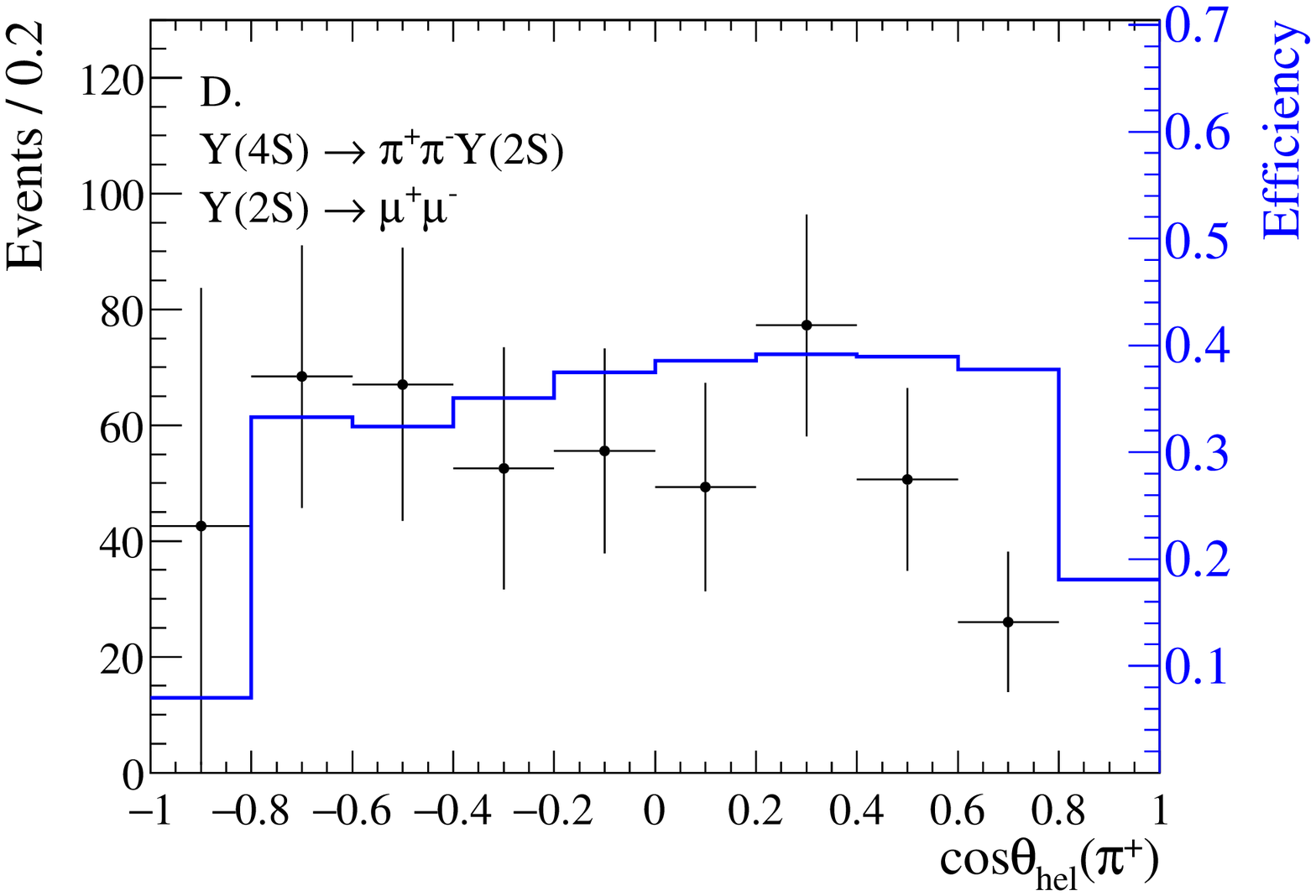}
\caption{Efficiency-corrected distributions of the helicity angle of the positive pion ($\cos\theta_{\rm hel}(\pi^+)$) for the signal component unfolded from the data distributions with the \textit{$_s$}${\cal P}$\textit{lot} technique~\cite{ref:splot} in the $\Upsilon(2S,3S,4S)\to\pi^+\pi^-\Upsilon(1S,2S)$ candidates. The values of the selection efficiency in each bin are shown in the blue histogram (right axis).}\label{fig:splot_hel}
\end{figure*}

\vspace{5mm}
For the dipion transitions, additional information can be obtained by the study of the dipion system invariant mass $M(\pi^+\pi^-)$, and of the angular distribution of the pions. The relevant distributions are shown in Figs.~\ref{fig:splot_mass} and~\ref{fig:splot_hel}, and are obtained by unfolding the signal component in the data distribution either in the $M(\pi^+\pi^-)$ or in the $\cos\theta_{\rm hel}(\pi^+)$ variable, according to the \textit{$_s$}${\cal P}$\textit{lot} technique described in~\cite{ref:splot}.
 
The invariant mass distributions for the $\Upsilon(4S)\to\pi^+\pi^-\Upsilon(2S)$ and $\Upsilon(3S)\to\pi^+\pi^-\Upsilon(1S)$ transitions show a doubly-peaked structure, with a clear enhancement near the dipion invariant mass threshold, that cannot be consistent with a pure phase-space description, as already shown by BaBar~\cite{ref:BaBar4S-previous} and CLEO~\cite{ref:CLEOdipion}.

The invariant mass distribution for the $\Upsilon(4S)\to\pi^+\pi^-\Upsilon(1S)$ transition shows an enhancement followed by a clear dip around 1~GeV/$c^2$, likely due to a contribution from the $f_{\rm 0}$(980) scalar meson and its interference with a non-resonant model. A similar pattern has been observed in the dipion transitions from $\Upsilon$ resonances above the $B\bar{B}$ threshold~\cite{ref:belle5S,ref:belleScan}, and has been recently predicted by theory~\cite{ref:Chen2016}.

In order to verify the $f_0(980)$ hypothesis, a $\chi^2$-fit is performed to the efficiency-corrected $M(\pi^+\pi^-)$ distribution for the signal events selected for the transition $\Upsilon(4S)\to\pi^+\pi^-\Upsilon(1S)$, as shown in Fig.~\ref{fig:splot_mass}(C).

The amplitude model is constructed either with a non-resonant component only, or by adding to this a contribution from $\Upsilon(1S)f_0(980)$. Each component $j$ is added to the model as a term of the form $A_{\rm{j}}e^{i\delta_{j}}$, where $A_{j}$ and $\delta_{j}$ are the amplitude and phase of the component, respectively.
The non-resonant component is parameterized by a first-order polynomial in $M^2(\pi^+\pi^-)$, as suggested in~\cite{ref:Voloshin2006,ref:Voloshin2007}: 
\begin{equation}
{\cal A}_{\rm{NR}}(M^2(\pi^+\pi^-)) = A_{\rm{NR}}^0e^{i\delta_{\rm{NR}}^0} + A_{\rm{NR}}^1e^{i\delta_{\rm{NR}}^1}M^2(\pi^+\pi^-). \nonumber
\end{equation}
Being sensitive to the relative phases and amplitudes only, the amplitude and phase of the lowest-degree term of the non-resonant model are arbitrarily fixed to 1 and 0, respectively.
In the $f_0(980)$ contribution:
\begin{equation}
{\cal A}_{f_0}(M^2(\pi^+\pi^-)) = A_{f_0}e^{i\delta_{f_0}}a_{f_0}(M^2(\pi^+\pi^-)),\nonumber
\end{equation}
$a_{f_0}$ is parameterized as a Flatt\'e function~\cite{ref:Flatte} with mass and coupling constants fixed to the values measured in the analysis of $B^+\to K^+\pi^+\pi^-$ events~\cite{ref:Garmash2005}, and used in~\cite{ref:belle5S}, $M(f_0(980))=950$ MeV/$c^2$, $g_{\pi\pi}=0.23$ and $g_{KK}=0.73$.  
An additional resonant contribution from $\Upsilon(1S)f_2(1270)$, with the $f_2(1270)$ component described by a relativistic Breit-Wigner function with mass and width fixed to the world average values~\cite{ref:PDG2016}, has been incoherently added to the amplitude model, but does not lead to an improvement in the description of data. 

\begin{figure}[!h]
\includegraphics[width=0.44\textwidth]{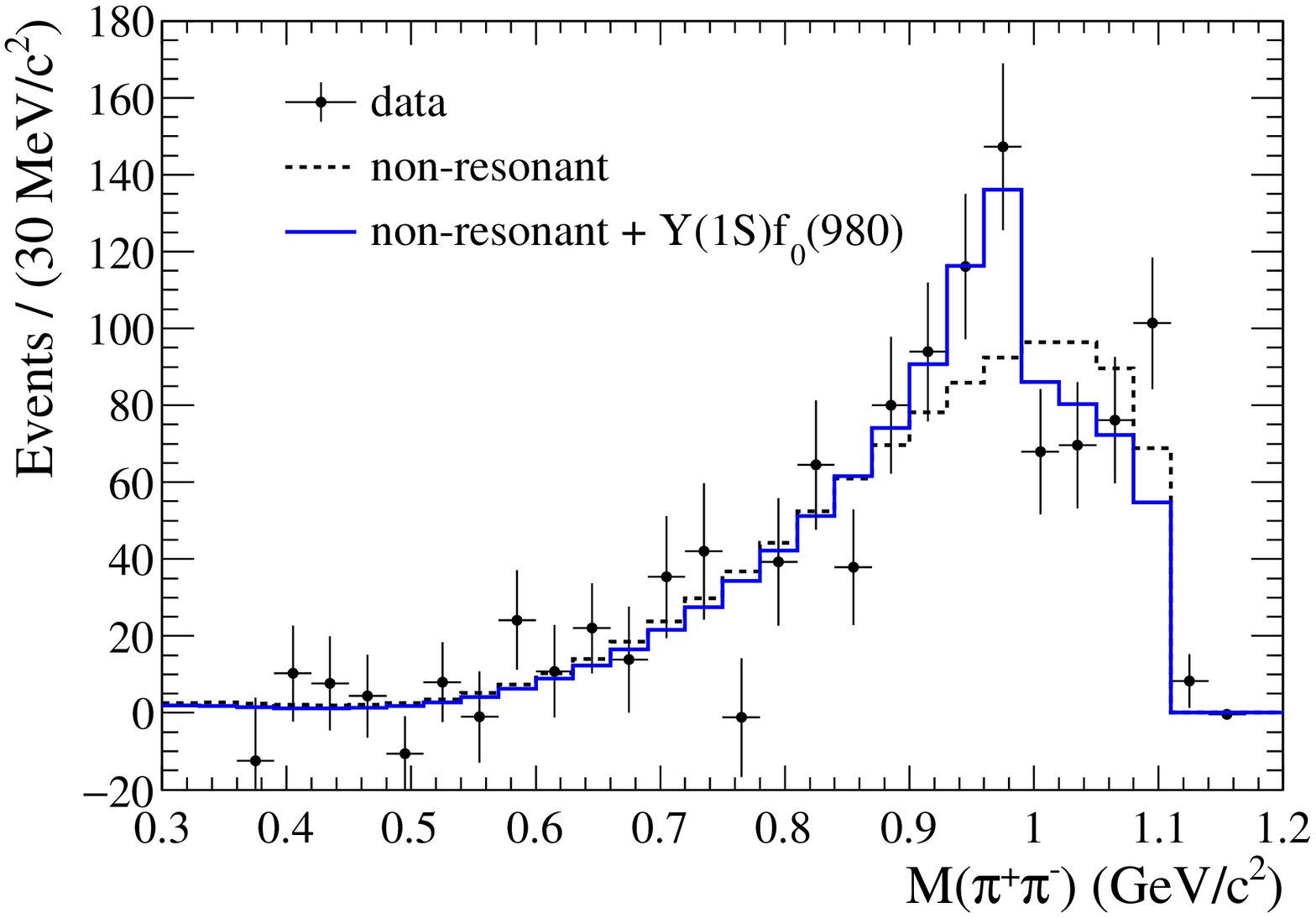}
\caption{Fit to the efficiency-corrected distribution of  $M(\pi^+\pi^-)$  for the signal component unfolded from the data distribution with the \textit{$_s$}${\cal P}$\textit{lot} technique~\cite{ref:splot}, in the $\Upsilon(4S)\to\pi^+\pi^-\Upsilon(1S)$ candidates (black points). The models used for the fit are: non-resonant model (black dashed line), and non-resonant $+\Upsilon(1S)f_0(980)$ model (blue solid line).}\label{fig:DP_fit}
\end{figure}

The fit results for the non-resonant only and the non-resonant $+~\Upsilon(1S)f_0(980)$ models are shown in Fig.~\ref{fig:DP_fit} and summarized in Table~\ref{tab:DP_fit}.
The model that includes the contribution from the $f_0(980)$ meson is preferred by the data, with a statistical significance of 2.8$\sigma$ according to Wilks' theorem~\cite{ref:Wilks}.

\begin{table}[]
\caption{Fit results for the amplitudes and phases of each component in the two models, obtained on the $\Upsilon(4S)\to \pi^+ \pi^- \Upsilon(1S)$ candidates selected in data. The value of the $\chi^2$ obtained in each fit is also shown, along with the number of degrees of freedom ($ndof$) and the corresponding $p$-value.}
\label{tab:DP_fit}
\begin{tabular}
 {@{\hspace{0.5cm}}l@{\hspace{0.5cm}}  @{\hspace{0.5cm}}c@{\hspace{0.5cm}} @{\hspace{0.5cm}}c@{\hspace{0.5cm}}}
\hline \hline
Parameter & Non-resonant & $+\Upsilon(1S)f_0(980)$\\
\hline
$A^0_{NR}$       & 1 (fixed)       & 1 (fixed)\\
$\delta^0_{NR}$ & 0 (fixed)       & 0 (fixed)\\
$A^1_{NR}$       & $4.63\pm0.23$       & $4.21\pm0.36$\\
$\delta^1_{NR}$ & $3.56\pm0.30$      & $-2.74\pm0.42$\\
$A_{f_0}$           &  -                  & $-0.14\pm0.04$\\
$\delta_{f_0}$     & -                  & $-0.28\pm0.47$\\
\hline \hline
$\chi^2$             & 41.9             & 31.4\\
$ndof$               & 26                & 24\\
$p$-value          & 0.025           & 0.142\\
\hline\hline
\end{tabular}
\end{table}

The analysis therefore shows indications for an $f_0(980)$ contribution.
A higher-statistics data sample, to be collected at the upcoming Belle II experiment, will allow for more precise studies.

\vspace{5mm}

We thank the KEKB group for the excellent operation of the
accelerator; the KEK cryogenics group for the efficient
operation of the solenoid; and the KEK computer group,
the National Institute of Informatics, and the 
PNNL/EMSL computing group for valuable computing
and SINET5 network support.  We acknowledge support from
the Ministry of Education, Culture, Sports, Science, and
Technology (MEXT) of Japan, the Japan Society for the 
Promotion of Science (JSPS), and the Tau-Lepton Physics 
Research Center of Nagoya University; 
the Australian Research Council;
Austrian Science Fund under Grant No.~P 26794-N20;
the National Natural Science Foundation of China under Contracts 
No.~10575109, No.~10775142, No.~10875115, No.~11175187, No.~11475187, 
No.~11521505 and No.~11575017;
the Chinese Academy of Science Center for Excellence in Particle Physics; 
the Ministry of Education, Youth and Sports of the Czech
Republic under Contract No.~LTT17020;
the Carl Zeiss Foundation, the Deutsche Forschungsgemeinschaft, the
Excellence Cluster Universe, and the VolkswagenStiftung;
the Department of Science and Technology of India; 
the Istituto Nazionale di Fisica Nucleare of Italy; 
the WCU program of the Ministry of Education, National Research Foundation (NRF)
of Korea Grants No.~2011-0029457, No.~2012-0008143,
No.~2014R1A2A2A01005286,
No.~2014R1A2A2A01002734, No.~2015R1A2A2A01003280,
No.~2015H1A2A1033649, No.~2016R1D1A1B01010135, No.~2016K1A3A7A09005603, No.~2016K1A3A7A09005604, No.~2016R1D1A1B02012900,
No.~2016K1A3A7A09005606, No.~NRF-2013K1A3A7A06056592;
the Brain Korea 21-Plus program, Radiation Science Research Institute, Foreign Large-size Research Facility Application Supporting project and the Global Science Experimental Data Hub Center of the Korea Institute of Science and Technology Information;
the Polish Ministry of Science and Higher Education and 
the National Science Center;
the Ministry of Education and Science of the Russian Federation and
the Russian Foundation for Basic Research;
the Slovenian Research Agency;
Ikerbasque, Basque Foundation for Science and
MINECO (Juan de la Cierva), Spain;
the Swiss National Science Foundation; 
the Ministry of Education and the Ministry of Science and Technology of Taiwan;
and the U.S.\ Department of Energy and the National Science Foundation.

\end{document}